\documentclass[%
 reprint,
 superscriptaddress,
 showpacs,
 amsmath,amssymb,
 prx,
 twocolumn
]{revtex4-2}
\usepackage{setspace}
\usepackage{verbatim}
\usepackage{lmodern}
\usepackage{siunitx}
\usepackage{etoolbox}
\usepackage{booktabs}
\usepackage{tabularx}
\usepackage{makecell}
\usepackage{graphicx}% Include figure files
\usepackage{dcolumn}% Align table columns on decimal point
\usepackage{bm}% bold math
\usepackage{bbm}
\usepackage{bbold}
\usepackage{physics}
\usepackage{enumerate}
\usepackage[dvipsnames]{xcolor}
\usepackage{times}
\definecolor{darkblue}{HTML}{023e8a}
\definecolor{darkred}{HTML}{780000}
\definecolor{darkgreen}{HTML}{005f73}
\definecolor{darkcyan}{HTML}{2a9d8f}
\usepackage[colorlinks=true,
	    urlcolor=darkred,
	    citecolor=darkblue,
	    linkcolor=darkred
            ]{hyperref}

\begin{document}
\title{
    %Topologically induced long-lived temporal interference in monitored quantum dynamics\\
  Temporal Interference from Topological Transitions in Monitored Quantum Dynamics
}

\author{Qingyuan Wang}
% \thanks{These authors contributed equally to this work.}
% \email{qingwqy@gmail.com}
\altaffiliation{These authors contributed equally\\
qingwqy@gmail.com}
\affiliation{Department of Physics, 
Institute of Nanotechnology and Advanced Materials, 
Bar-Ilan University, Ramat-Gan 52900, Israel}

\author{Ruoyu Yin}
%\collaboration{Both authors contributed equally to this work.}
% \thanks{These authors contributed equally to this work.}
\altaffiliation{These authors contributed equally\\
qingwqy@gmail.com}
%\email{yinruoy@biu.ac.il}
\affiliation{Department of Physics \#1, Graduate School of Science, Kyoto University, Kyoto 606-8502, Japan}
 
\author{Eli Barkai}
\affiliation{Department of Physics,   
Institute of Nanotechnology and Advanced Materials, 
Bar-Ilan University, Ramat-Gan 52900, Israel}

\begin{abstract}
% %We uncover a form of dark-state temporal interference in stroboscopically monitored quantum dynamics that treats the back action of measurements and the unitary dynamics.
% \newc{We uncover \textit{long-lived} temporal interference in stroboscopically monitored quantum dynamics, arising from the interplay of measurement backaction and unitary evolution.}
% Previous work established that the mean detected recurrence time is integer-quantized and given by a topological invariant, a winding number $w$.
% When measurement periods are at resonance with the system's timescales,
% the winding number can abruptly change.
% We focus on a generic quantum system and the transition $w\to w-2$,
% signified by the creation of two dark states in Hilbert space, whose corresponding modes are responsible for the interference pattern.
% Close to the transition an extremely slow decay of the amplitude of first detection is found,
% superimposed by oscillations,
% in contrast to the monotonically exponential decay close to the case $w\to w-1$.
% We show how these oscillations are obtained from the symmetry of the system and find the conditions for optimal observations of the phenomenon.
% % Our results establish a direct link between graph topology and transient quantum dynamics,
% % providing a framework for spectral engineering in monitored quantum systems.

Quantum interference is conventionally associated with the coherent superposition of alternative paths. We uncover a temporal form of interference in repeatedly monitored quantum systems, appearing directly as fringes in the stochastic first-detection-time distribution. This temporal interference becomes remarkably long lived near a class of topological transitions at which a pair of dark states emerges in Hilbert space. Close to such transitions, the corresponding quasi-dark modes dominate the late-time dynamics and interfere coherently, producing pronounced temporal fringes whose decay becomes arbitrarily slow upon approaching the transition. We identify the symmetry conditions underlying this mechanism and relate the fringe frequency to the energy spectrum of the Hamiltonian. Our results reveal how the interplay of quantum superposition, topological transitions, and dark-state formation generates long-lived temporal interference in monitored quantum dynamics.

%Quantum interference is conventionally associated with the coherent superposition of alternative paths. We uncover a temporal form of interference in repeatedly monitored quantum systems, appearing directly as fringes in the stochastic first-detection-time distribution. 
% The effect becomes remarkably long lived near a class of topological transitions, where two modes of the monitored dynamics approach the dark-state limit and form coherent temporal channels. 
%This temporal interference becomes remarkably long lived near a class of topological transitions associated with the creation of a pair of dark states in Hilbert space. Close to such transitions, the corresponding quasi-dark modes dominate the late-time dynamics and interfere coherently, producing pronounced temporal fringes whose decay becomes arbitrarily slow upon approaching the transitions.
%Their superposition produces pronounced oscillations whose decay becomes arbitrarily slow upon approaching the transition. 
%We identify the symmetry conditions underlying this mechanism and relate the fringe frequency to the energy spectrum of the Hamiltonian. 
%Our results show how superposition, time-topology and dark state creation controls long lived temporal interference 
%.
%Our results show that topology can govern not only stationary or quantized observables, but also coherent transient dynamics generated by quantum measurement.}

\end{abstract}

\maketitle

\section{Introduction}
In quantum mechanics, observables are represented by operators, while measurement outcomes are random variables. A familiar example is the wave function $\psi(x,t)$, the probability amplitude whose evolution is governed by the Schr\"odinger equation with Hamiltonian $H$. Interference arises when the amplitude is composed of several contributions. For example,
\[
\psi(x,t)=A_1(x) e^{i\Phi_1(x)}+A_2(x) e^{i\Phi_2(x)},
\]
where the amplitudes $A_1(x)$ and $A_2(x)$ are real.
If one amplitude dominates, interference is weak. The case $A_1(x)=A_2(x)=A(x)$ is of special interest, since the interference pattern is then most pronounced, as in the double-slit experiment \cite{Gerry_Knight_2004}. In that case, the corresponding  probability is
\[
|\psi(x,t)|^2=4A^2(x)\cos^2\left(\frac{\Delta\Phi(x)}{2}\right),
\]
where $\Delta\Phi(x)=\Phi_1(x)-\Phi_2(x)$. 
Here $A^2(x)$ is a slowly varying function of $x$ compared with the length scale on which $\Delta\Phi(x)$ oscillates, so the interference fringes are modulated by a slowly varying envelope.

Recently, a different type of quantum amplitude, describing the random hitting time of monitored quantum processes, has attracted considerable attention \cite{Muga2000,Hiroshi2003,kempe2005,Krovi2006a,Krovi2008,Gruenbaum2013,Dhar2015a,dhar2015,Friedman2017a,Kulkarni2023,Bravo2025,EricLutz2025,liu2025_large_deviations, Seth2026,Paul2026,Vecchio2026,Stefanak2026b,Imparato2026}. We consider a finite-dimensional quantum system subjected to a stroboscopic protocol: unitary evolution $U(\tau)=e^{-iH\tau}$ (with $\hbar=1$) followed by a projective measurement that asks whether the system is in a target state $\ket{x_T}$. 
This procedure is repeated, with a measurement performed every sampling time $\tau$.
The outcome is a random detection step $n$, namely the first measurement at which the system is found in $\ket{x_T}$. Depending on the initial state, this setting describes either a first-hitting or a recurrence problem \cite{kempe2005,Krovi2006a,Krovi2006,Krovi2008,Gruenbaum2013,Dhar2015a,dhar2015, Friedman2017a,yin2019,dubey2021quantum,Das2022,Zhenbo2023,yajing2023,Xiaoxiao2024,Roy2025,Matteo2025,Shukla2024accelerated,Stefanak2026}.
In the classical setting, the analogous questions are the first-passage and first-return problems \cite{Redner2001, Hanggi1990,Bray2013,Metzler2014book,Benichou2014}. The monitored dynamics is nonunitary due to measurement backaction. Since the limit $\tau\to 0$ is singular due to the quantum Zeno effect \cite{Misra1977, Lahiri2019, Thiel2020J,Liu2023Feng}, $\tau$ is not taken too small and instead serves as a control parameter. 
The periodic sampling also gives the dynamics a Floquet-like stroboscopic structure \cite{Bukov2014,Oka2019,Rudner2020,Mori2023}.
Quantum hitting time statistics exhibit many other novel features compared with corresponding classical processes, for example, dark state physics \cite{Krovi2006,Thiel2020D} and the quantum advantage related to quantum search \cite{kempe2005,Krovi2006a,Magniez2008,Ruoyu2023,Guimaraes2026}.
More broadly, the interplay of unitary evolution and measurement backaction places first-detection processes within the wider field of monitored quantum dynamics \cite{Fisher2018,Skinner2019,Chan2019,Hoke2023,Fisher2023}.

Let $\phi(n,x_T)$ denote the amplitude for first detection in the target state at time $n\tau$. The corresponding first-detection probability is $F_n=|\phi(n,x_T)|^2$ \cite{kempe2005,Krovi2006a,Krovi2006,Krovi2008,Gruenbaum2013,Dhar2015a,dhar2015,Friedman2017a}. Our starting point is the observation that $\phi(n,x_T)$ itself may display interference, namely
\[
\phi(n,x_T)=\tilde{A}_1(n) e^{i\tilde{\Phi}_1(n)}+\tilde{A}_2(n) e^{i\tilde{\Phi}_2(n)}.
\]
As in the spatial case, interference is strongest when $\tilde{A}_1(n)=\tilde{A}_2(n)=\tilde{A}(n)$, for which
\begin{equation}\label{eq:F_n_1}
F_n=4\tilde{A}^2(n)\cos^2\left(\frac{\Delta\tilde{\Phi}(n)}{2}\right),
\end{equation}
with $\Delta\tilde{\Phi}(n)=\tilde{\Phi}_1(n)-\tilde{\Phi}_2 (n)$.
Our setting differs from temporal double-slit schemes \cite{Szriftgiser1996,Lindner2005,Tirole2023}, where  slits are windows in time and the fringes typically appear in energy or frequency spectra. Here, instead, the interference appears directly in the first-detection probability as a function of the \textit{detection time}, and originates from a pair of quasi-dark modes, as discussed below.
Our goal is to uncover the physics behind this structure: when can the condition $\tilde{A}_1(n)=\tilde{A}_2(n)$ be realized, and how do the parameters on the right-hand side depend on $\tau$, $n$, and the Hamiltonian $H$? We show that these questions are tied to special choices of the sampling time, to symmetries of $H$ and recurrence, and ultimately to topological effects and the emergence of dark states in Hilbert space.

\begin{figure*}[t]
    \centering
    \includegraphics[width=0.95\textwidth]{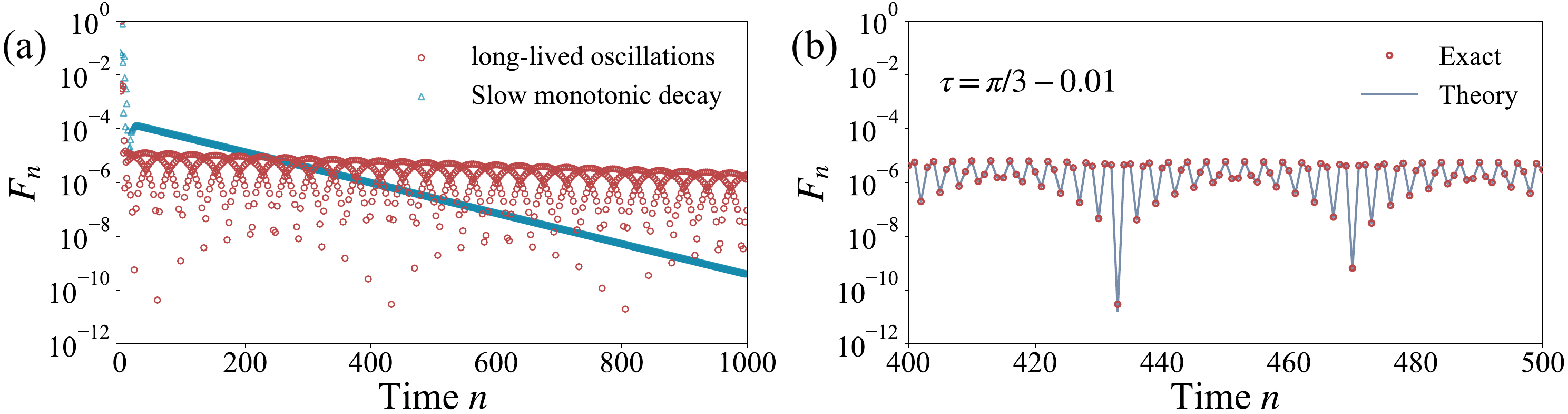}
    \caption{
    (a)~First-detection probability $F_n$ versus measurement number $n$ for a monitored quantum walk on a $d=4$ hypercube, shown for two sampling intervals $\tau$. 
    %measured in units where the hopping amplitude is unity; see details in End Matter.} 
    Blue triangles, $\tau=\pi/4-0.01$, display a slow monotonic decay, while red circles, $\tau=\pi/3-0.01$, exhibit long-lived oscillations.
(b)~Close-up for $\tau=\pi/3-0.01$, comparing exact numerical results from Eq.~\eqref{eq:phi_n} with the asymptotic prediction of Eq.~\eqref{eq:interference1}. The latter is constructed from the two eigenvalues of ${\cal S}$ closest to the unit circle, with the phase $\beta$ defined in Eq.~\eqref{eq:sm_beta}. The theory captures both the slow decay of the envelope and the oscillation period at large $n$.
    }
    \label{fig:fig_protocol1}
\end{figure*}

\section{Temporal interference} 
Our first claim is that the recurrence problem gives rise to long-lived interference patterns in time. 
 Specifically, this implies that $\tilde{A}(n)$ is decaying very  slowly.
 For the recurrence problem, the process starts and ends at the target state $\ket{x_T}$, so a detected recurrence effectively closes a loop in Hilbert space. 
 Consider a finite-dimensional system initially prepared in $\ket{x_T}$ and evolving under
$
U(\tau)=e^{-iH\tau}.
$
After each interval $\tau$, the system is measured: if it is found in $\ket{x_T}$, the process stops; otherwise the component along $\ket{x_T}$ is projected out and the evolution continues. The amplitude for first detection at step $n$ is \cite{Gruenbaum2013, Friedman2017a}
\begin{equation}\label{eq:phi_n}
\phi(n,x_T)=\bra{x_T}U(\tau)\mathcal{S}^{\,n-1}\ket{x_T},
\end{equation}
where
$
\mathcal{S}=(\mathbb{1}-\ket{x_T}\bra{x_T})U(\tau)
$
is the survival operator.
%, namely the evolution conditioned on no detection. 
This expression describes $n-1$ failed detection attempts followed by success on the $n$th attempt. For finite-dimensional Hilbert spaces, the first-detection probabilities are normalized, $\sum_{n=1}^\infty F_n=1$ \cite{Gruenbaum2013}.
% [The Gr\"unbaum quantization result <n>=w and the resonant transitions were moved to the section "Topological transitions and dark states" below.]

The eigenvalues of the survival operator ${\cal S}$ provide the key to the mechanism behind temporal interference. We denote them by $\{\xi\}$ and separate them into two classes: eigenvalues on the unit circle, $|\xi|=1$, and eigenvalues strictly inside the unit disk, $|\xi|<1$. Since ${\cal S}$ depends on the sampling time $\tau$, its spectrum changes as $\tau$ is varied. In particular, an eigenvalue may cross from the unit disk to the unit circle; as shown below, such crossings correspond to topological transitions of the monitored dynamics.

These two spectral classes have a clear physical meaning. Eigenstates with $|\xi|<1$ are bright states, which are eventually detected with probability unity. Eigenstates with $|\xi|=1$ are dark states, which remain undetected \cite{Thiel2020D,Liu2022a}. The Hilbert space is therefore fragmented into bright and dark subspaces, and tuning the sampling time $\tau$ can change their dimensions. Although the target state $\ket{x_T}$ lies in the bright subspace, the formation of dark states elsewhere in Hilbert space is crucial for temporal interference. As we show below, temporal interference in the recurrence problem becomes long lived near points where the dimension of the dark subspace changes.

The first-detection amplitude admits the spectral decomposition
\begin{equation}\label{eq:phi_n_decom}
\phi(n,x_T) = \sum_{k=0}^{w-1} r_k e^{i\beta_k}\xi_k^n ,
% \label{eqphi}
\end{equation}
where $\{\xi_k\}$ are the eigenvalues of ${\cal S}$ inside the unit disk, $|\xi_k|<1$, and $w$ is the number of such eigenvalues (including one trivial eigenvalue $\xi=0$).
The coefficients $r_k e^{i\beta_k}$ are the corresponding spectral weights, given explicitly in Appendix Eq.~\eqref{eq:sm_coefficients}. At a topological transition, $w$ changes \cite{Gruenbaum2013}, and hence the number of terms contributing to Eq.~\eqref{eq:phi_n_decom} changes as well.

For large $n$, the behavior of $\phi(n,x_T)$ is controlled by the eigenvalues of ${\cal S}$ closest to the unit circle. If a single eigenvalue dominates, then the first-detection probability $F_n=|\phi(n,x_T)|^2$ decays exponentially at long times, and no temporal interference survives. 
This behavior is illustrated by the blue triangles in Fig.~\ref{fig:fig_protocol1}(a) for a quantum walk on a four-dimensional hypercube, with hopping amplitude equal to unity (see Appendix for details).
The exponential tail itself is a general feature whenever the long-time dynamics is governed by a single leading eigenvalue.
%The quantum walks studied in this work are tight-binding models with hopping amplitude $\gamma$, which sets the energy scale of the problem; we set $\gamma=1$, so that the sampling interval $\tau$ is measured in units of $\hbar/\gamma$. The models are defined in the End Matter.

In contrast, when a pair of eigenvalues, denoted $\xi_1$ and $\xi_2$, approaches the unit circle simultaneously, the long-time behavior is controlled by two long-lived eigenmodes of $\mathcal{S}$, with $|\xi_1|=|\xi_2|$ just below unity. These modes are close to becoming dark, and their coexistence gives rise to temporal interference.
To demonstrate this effect and sharpen the underlying questions, we evaluate $F_n$ from Eq.~\eqref{eq:phi_n} for a monitored quantum walk on a four-dimensional hypercube, choosing the sampling time $\tau=\pi/3-0.01$.
As shown by the red circles in Fig.~\ref{fig:fig_protocol1}(a), $F_n$ exhibits long-lived oscillations after a short transient. This behavior is striking: $F_n$ is a normalized first-detection distribution, yet the decay of $F_n$ is hardly visible. The key questions are therefore what sets the oscillation frequency, what controls the envelope and slow relaxation, and how these features are connected to the emergence of dark states near topological transitions.

As mentioned, in the long-time limit, the faster modes are exponentially suppressed, and the
behavior of $\phi(n,x_T)$ is controlled by the pair $\xi_1$ and $\xi_2$.
Writing $\xi_1 = |\xi_1| e^{i\theta_1}$ and $\xi_2 = |\xi_2| e^{i\theta_2}$, we note that the interference can be sustained in the large-$n$ limit only if 
$
|\xi_1| = |\xi_2| = |\xi| 
$
and $|\xi|$ is close to unity implying a slow decay. 
As we show below, this matching is related to bipartite symmetry of the underlying
Hamiltonian, and thus it is a feature found for wide classes of quantum dynamics.
% [the bipartite-symmetry sentence that stood here was moved, verbatim, to the topology section below]
Collecting the two modes, we obtain
\begin{equation}
\phi(n,x_T) \sim |\xi|^n \left( r_1 e^{i\theta_1 n + i \beta_1}
+ r_2 e^{i\theta_2 n + i\beta_2} \right).
\label{eq:two_mode_phi}
\end{equation}
This form shows that a weak overall decay is accompanied by oscillations set by
the phase difference $\theta_1-\theta_2$. A pronounced interference pattern
requires comparable weights, $r_1=r_2$, which
\mbox{\textit{a priori}} appears unlikely. Remarkably, one of our key results is that this is precisely what
occurs close to the topological transitions at which a pair of eigenvalues approaches the unit circle.
% [deleted: "the $w \to w - 2$ transition" -- recovered below]

% [QW: ``companion paper'' -> Supplemental Material throughout, since we no longer plan a joint submission.]
% In a companion paper \cite{Wang2026_compL} we present a detailed derivation and obtain
In Appendix we present a detailed derivation and obtain
\begin{equation}
    %\boxed{
    F_n \sim 
     \underbrace{
    4 (1-|\xi|^2)^2 |\xi|^{2n}
     }_{\text{Envelope Decay}}
     \times
     \underbrace{
    \cos^2
    \left[ 
    {n\over 2} (\theta_1-\theta_2) + \beta 
    \right]
    }_{\text{Interference Fringes}}.
    %}
    \label{eq:interference1}
\end{equation}
Recall that $|\xi|$ is controlled by the sampling time $\tau$ (see below). When $\tau$ is tuned close to such a transition, $|\xi|\simeq 1$, and the oscillations arising from temporal interference become remarkably long lived, since the envelope decays only slowly. Precisely at the transition, where $|\xi|=1$, the effect disappears due to the formation of dark states. The oscillation frequency is set by the phase difference $\Delta \theta=\theta_1-\theta_2$ of the relevant eigenvalues, so when these phases are close the oscillation period is long. In contrast, the phase $\beta$ entering Eq.~\eqref{eq:interference1} depends on all eigenvalues of $\mathcal{S}$ and does not admit such a simple description, although an explicit expression is given in Appendix Eq.~\eqref{eq:sm_beta}. The prediction of Eq.~\eqref{eq:interference1} is compared with the exact numerics in Fig.~\ref{fig:fig_protocol1}(b), showing excellent agreement at large $n$.
% [in the three sentences above, "the transition" / "the $w\to w-2$ transition" was replaced by "resonance", since the transitions are defined below]
%Where this special pair of eigenvalues comes from is, as we show below, dictated by the spectrum of the underlying Hamiltonian.

\begin{figure}[t]
    \centering
    \includegraphics[width=0.48\textwidth]{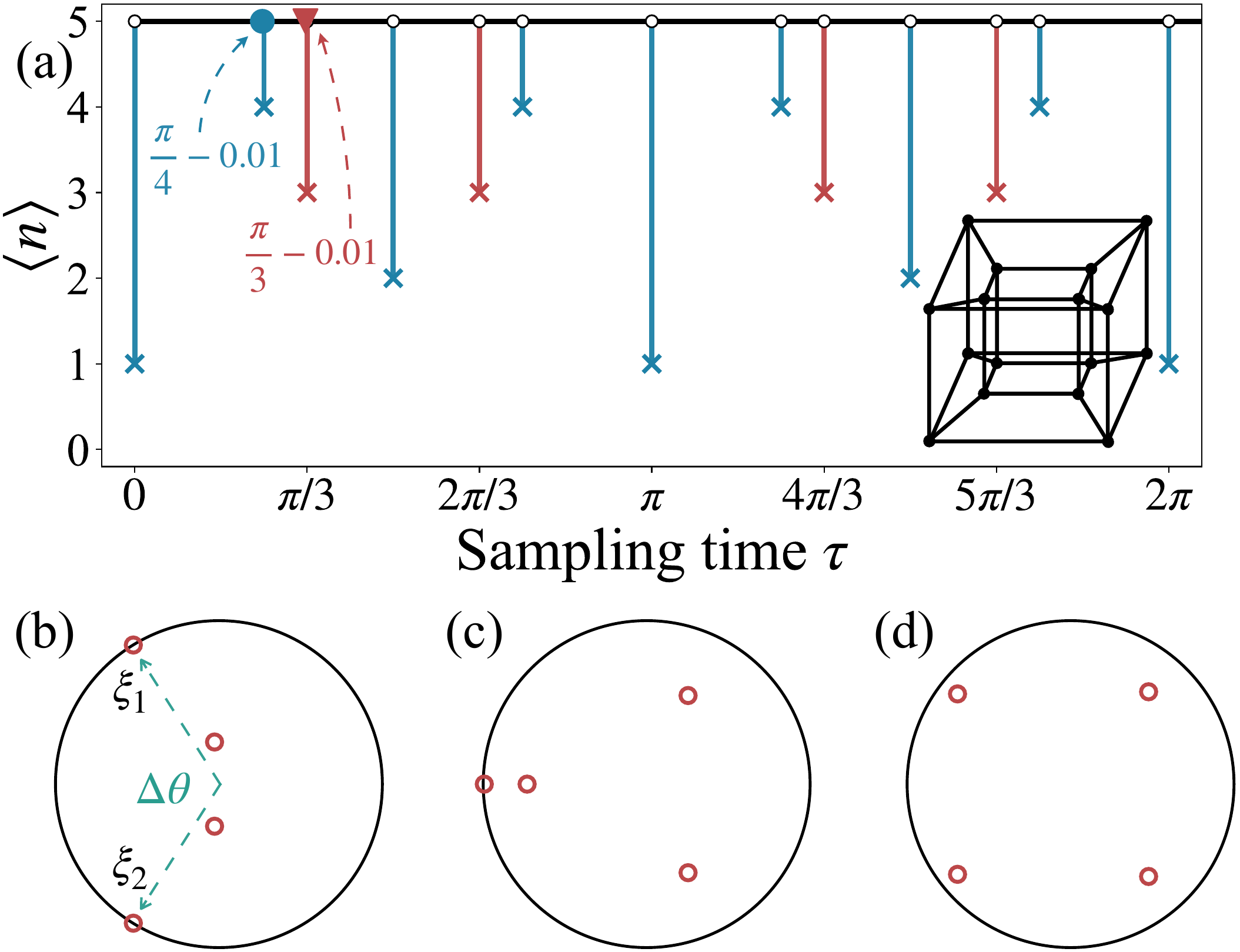}
    \caption{
    (a)~Mean recurrence time $\langle n \rangle$, or equivalently the winding number $w$, versus sampling interval $\tau$ for a $d=4$ hypercube (inset). Here the typical value $w=5$ reflects five distinct energy levels and undergoes discrete transitions at resonances. 
    %The typical winding number $w=5$ reflects five distinct energy levels and undergoes discrete transitions at resonances.
    (b)~Eigenvalues of the survival operator $\mathcal{S}$ in the unit disk, close to the $w=5\to3$ transition ($\tau=\pi/3-0.01$, red triangle in (a)). 
    Two dominant eigenvalues $\xi_1,\xi_2$ approach the unit circle with phase splitting $\Delta\theta$ and govern the long-lived oscillations since $|\xi_1| = |\xi_2| \simeq 1$; other eigenvalues of $\mathcal{S}$ remain well inside the unit circle and hence do not contribute.
    (c)~Near the $w=5\to4$ transition ($\tau=\pi/4-0.01$, blue dot in (a)), a single eigenvalue approaches the unit circle, producing the slow monotonic decay of $F_n$.
    (d)~Far from topological transitions ($\tau=\pi/5$), all eigenvalues lie well inside the unit circle.
    }
    \label{fig:Fn_4hypercube}
\end{figure}

\section{Topological transitions}
The remaining question is when, and why, a pair of eigenvalues of ${\cal S}$ approaches the unit circle. The answer is tied to the topology of the monitored dynamics.
A key result due to Gr\"unbaum \textit{et al.} \cite{Gruenbaum2013} is that the mean recurrence time is quantized. A quantum extension of Kac's recurrence lemma \cite{kac1947} gives
\begin{equation}
\langle n\rangle=w,
\end{equation}
where $w$ is an integer winding number associated with the generating function of the problem. In the spectral representation used here, $w-1$ also denotes the number of non-vanishing terms in the spectral sum, Eq.~\eqref{eq:phi_n_decom}.
As shown in previous works \cite{Gruenbaum2013,yin2024restart}, $w$, and therefore $\langle n\rangle$, can jump discontinuously at special sampling times, while remaining integer-valued. These transitions occur when phase factors of the unitary operator $U(\tau)$ become degenerate. Writing $U(\tau)\ket{E_k}=e^{-iE_k\tau}\ket{E_k}$, the degeneracy condition is
$
e^{-iE_1\tau}=e^{-iE_2\tau},
$
with $E_1$ and $E_2$ any two distinct energies.
Varying $\tau$ therefore tunes the system through resonance conditions, where phase matching between energy levels produces abrupt changes in the winding number and hence of the observed $\langle n \rangle$.

A key result is that long-lived temporal interference appears close to transitions in which the winding number changes by two, $w\to w-2$. In general, $w$ is the number of distinct phase factors $e^{-iE_k\tau}$ for which $\langle E_k|x_T\rangle\neq 0$. It can therefore be controlled by the sampling time $\tau$, as in the examples considered here, but also by other external control parameters \cite{Gruenbaum2013,Friedman2017a}.
The special sampling times can be identified by plotting the mean recurrence time $\langle n\rangle$, obtained either theoretically or experimentally, as a function of $\tau$. Topological transitions then appear as discontinuous jumps of $\langle n\rangle=w$ that are widened in experiments \cite{Wang2023}. 
Once a transition of the type $w\to w-2$ is located, the sampling time can be tuned close to this point in order to observe long-lived interference in the first-detection distribution $F_n$.

For example, consider the monitored quantum walk on a four-dimensional hypercube, shown in the inset of Fig.~\ref{fig:Fn_4hypercube}(a). In this case the generic value $w=5$ equals the number of distinct energy levels coupled to the target state, which is a node of the graph.
In Fig.~\ref{fig:Fn_4hypercube}(a), we plot $\langle n\rangle$ versus $\tau$ for this model. The topological transitions are directly visible as jumps in $\langle n\rangle$, for example from $5$ to $4$ and from $5$ to $3$. In the former case, $F_n$ decays exponentially, but slowly, whereas in the latter case it exhibits a long-lived oscillatory pattern.

Indeed, the value $\tau=\pi/3-0.01$ used in Fig.~\ref{fig:fig_protocol1}(b) lies close to the $5\to3$ transition of the hypercube, marked by the red triangle in Fig.~\ref{fig:Fn_4hypercube}(a), and therefore displays the sought-after interference effect. As mentioned above, the eigenvalues of ${\cal S}$ inside the unit disk determine whether the relaxation of $F_n$ is fast, slow, or oscillatory. 
%For the hypercube example, we focus on the special sampling times $\tau=k\pi/3$ with $k=1,2,4,5$ [BUT I SEE ONLY 3 EXAMPLES, ALSO YOU ARE USE $k \pi/3 - \eps$ SO YOU WORK IN VICINITY OF THESE TAUs], highlighted in Fig.~\ref{fig:Fn_4hypercube}(a). 
In Fig.~\ref{fig:Fn_4hypercube}(b), we plot the eigenvalues of ${\cal S}$ and observe that a pair of eigenvalues approaches the unit circle. This signals a $w\to w-2$ transition and gives rise to long-lived oscillations of $F_n$, with a frequency determined by the phases of these eigenvalues, as predicted by Eq.~\eqref{eq:interference1}.
For other choices of $\tau$, see Fig.~\ref{fig:Fn_4hypercube}(c)(d). In these cases, either all eigenvalues remain well inside the unit disk, leading to fast relaxation, or only a single eigenvalue approaches the unit circle. The latter corresponds to the formation of a single dark state and a $w\to w-1$ transition, yielding a simple but slow exponential decay of $F_n$ \cite{yin2024restart,YinWang2025}; see also Appendix.

\section{Hamiltonian origin of temporal interference}
So far, we have described the emergence of long-lived temporal interference in terms of the spectrum of ${\cal S}$. This spectral description directly identifies the slowly decaying modes once they are present. To predict when such modes appear, however, it is useful to return to the underlying Hamiltonian.
The relevant condition is the near-merging of two independent pairs of unitary phase factors,
$e^{-iE_1\tau}\simeq e^{-iE_2\tau}$ and $e^{-iE_3\tau}\simeq e^{-iE_4\tau}$. 
%where $E_1,\dots,E_4$ again label any four distinct energies.
At exact resonance, the corresponding equalities hold, and each merged pair supports a dark state. For example, for the pair $(E_1,E_2)$ one may construct
$
    \ket{D_{12}} \propto
    \langle x_T|E_2\rangle \ket{E_1}
    -
    \langle x_T|E_1\rangle \ket{E_2},
$
which is orthogonal to the target state.
% This state is dark as it is never detected under the stroboscopic measurement protocol since the unitary only induces a global phase (it is also easy to see that it is an eigenstate of ${\cal S}$ with an eigenvalue on the unit circle).
This state is dark: the unitary evolution only changes its overall phase, so it is never detected; equivalently, it is an eigenstate of ${\cal S}$ with an eigenvalue on the unit circle.
A second dark state, $\ket{D_{34}}$, is formed in the same way from the pair $(E_3,E_4)$. Thus, at resonance, two eigenvalues of ${\cal S}$ lie on the unit circle, giving the topological transition $w\to w-2$.

The long-lived interference observed in $F_n$ occurs close to, but not exactly at, this condition. Then the two dark states are weakly coupled to the detector, and the corresponding eigenvalues $\xi_1$ and $\xi_2$ of ${\cal S}$ move slightly inside the unit circle. Assuming that the two nearly merged pairs remain well separated from all other unitary phase factors, these eigenvalues $\xi_1$ and $\xi_2$ inherit their phases from the underlying unitary dynamics, $\theta_1\simeq -E_{12}\tau,\, \theta_2\simeq -E_{34}\tau$,
% \begin{equation}
%     \xi_{1,2}=|\xi_{1,2}|e^{i\theta_{1,2}},
%     \qquad
%     \theta_1\simeq -E_{12}\tau,
%     \qquad
%     \theta_2\simeq -E_{34}\tau,
% \end{equation}
where
% $
%     E_{12}=(E_1+E_2)/2,\,
%     %\qquad
%     E_{34}=(E_3+E_4)/2.
% $
$E_{12}=(E_1+E_2)/2$ and $E_{34}=(E_3+E_4)/2$ are the centroids (mean energies) of the two nearly merged pairs.
% [QINGYUAN PLEASE CHECK and ADD YOUR NOTE: Since $U(\tau)=\exp(-iH\tau)$ and we write $\xi=|\xi|e^{i\theta}$, the phases are written here with a minus sign, modulo $2\pi$. If elsewhere we use the convention $\xi=|\xi|e^{-i\theta}$, the sign should be reversed.]
Consequently, the oscillation frequency of the temporal interference is set by the phase difference
\begin{equation}\label{eq:leading_theta}
    \Delta \theta=\theta_1-\theta_2
    \simeq (E_{34}-E_{12})\tau,
\end{equation}
% showing that the interference pattern is directly controlled by the energy structure of the Hamiltonian.
showing that, while the eigenvalues of $\mathcal{S}$ are the basic objects in these problems, it is the familiar energy levels of the Hamiltonian that determine the oscillations in Eq.~\eqref{eq:F_n_1}.
% The superscript $(0)$ denotes the leading order in the small detuning from exact resonance. The detuning also sets the lifetime of the interference: it determines how far $|\xi_{1,2}|$ falls below unity, and hence the slow decay of the envelope. The corrections beyond leading order, both to the phases $\theta_{1,2}$ and to the magnitudes $|\xi_{1,2}|$, are derived in SM using perturbation theory.

Importantly, the phase-matching condition arises naturally in systems with bipartite symmetry. In such systems the Hamiltonian spectrum is symmetric: for each energy $E$ there is a partner at $-E$, up to possible states with energy $E=0$. Hence the unitary phase factors appear in conjugate pairs, $e^{-iE\tau}$ and $e^{iE\tau}$, and suitable choices of $\tau$ can bring two independent pairs of phase factors close to merging. 
Bipartite symmetry thus provides a natural route to the $w\to w-2$ transition and to the resulting long-lived temporal interference. In the following we focus on bipartite systems and analyze two distinct scenarios in which the interference mechanism is realized.

\subsubsection*{Merging of pairs of phase factors}
The first case occurs when two pairs of unitary phase factors approach each other while remaining well separated from the rest, meaning that the phase difference in Eq.~\eqref{eq:leading_theta} is not small, as shown in Fig.~\ref{fig:Fn_spectral}(b). The long-time dynamics is then governed by these two resonant pairs.
% We use a perturbative approach, writing the phase difference as $\Delta\theta = \Delta\theta^{(0)} + \Delta\theta^{(1)}$, with the zeroth-order term given in Eq.~\eqref{eq:leading_theta}.
% We use a perturbative approach to find the leading-order small corrections to Eq.~\eqref{eq:leading_theta}. A detailed calculation, presented in SM, then yields the spectral response in a physically transparent form:
A perturbative calculation \cite{yin2019}, detailed in Appendix, yields the leading-order small corrections to Eq.~\eqref{eq:leading_theta} and the resulting spectral response in a physically transparent form:
\begin{equation}\label{eq:Fn_spectral}
    F_n \sim 4 \left[ \lambda (\widetilde{\Delta E \tau})^2 \right]^{2} e^{-\lambda (\widetilde{\Delta E \tau})^2 n} \cos^2 \left( \frac{n}{2} \Delta \theta + \beta \right),
\end{equation}
where the frequency is
\begin{equation}\label{eq:Fn_freq}
\Delta\theta = (E_{34}-E_{12})\tau + \delta, \qquad \delta = 2\eta\,\widetilde{\Delta E\tau},
\end{equation}
with
$\Delta E = E_{1}-E_{2}=E_{4}-E_{3}$ the associated energy splitting, equal for the two pairs since bipartite symmetry gives $E_{3}=-E_{1}$ and $E_{4}=-E_{2}$.
Throughout, the tilde denotes the reduced detuning $\widetilde{\Delta E \tau}\equiv\Delta E \tau\ (\mathrm{mod}\ 2\pi)\in(-\pi,\pi]$, which measures the departure from exact merging, and clearly $\widetilde{\Delta E\tau}=0$ at the topological transition.
The coefficients $\lambda$ and $\eta$ are determined by the spectral weights $P_{k} = \sum_m |\langle x_T | E_{k, m} \rangle|^2$ of the resonant levels, with $m$ accounting for any underlying degeneracy of energy level $E_k$. Specifically, the damping factor is $\lambda = P_{1} P_{2}/(P_{1} + P_{2})^3$ and the frequency shift is $\eta = (P_{1} - P_{2})/[2(P_{1} + P_{2})]$. 
For bipartite systems, the spectral symmetry gives $P_1=P_3$ and $P_2=P_4$, so the two resonant pairs have matching weights.
% In this regime, the oscillation reflects interference between two distinct ``temporal slits'' separated by a finite frequency.
% [QW: merged with the Eq.(8)/(9)->Eq.(1) interpretation into the single sentence below]
Physically, Eqs.~\eqref{eq:Fn_spectral} and~\eqref{eq:Fn_freq} make Eq.~\eqref{eq:F_n_1} explicit: the small detuning $\widetilde{\Delta E\tau}$ controls the slow envelope decay near the transition, while the finite frequency $\Delta\theta$, the relative phase between the two ``temporal slits,'' produces the fringes.
% Fig.~\ref{fig:Fn_spectral}(a) confirms that this spectral prediction (solid line) perfectly captures the exact dynamics (circles) of a tight-binding quantum walk on a seven-site chain, a distinct bipartite model that illustrates the generality of the mechanism,  as detailed in the End Matter.
% The same mechanism also underlies the hypercube discussed above, for which a detailed analysis is provided in the SM.
Fig.~\ref{fig:Fn_spectral}(a) confirms that this spectral prediction (solid line) captures the exact dynamics (circles) of a seven-site tight-binding chain, detailed in Appendix. The same mechanism underlies the hypercube discussed above, analyzed in Appendix.
\subsubsection*{Merging of three phase factors}
\begin{figure}[t]
    \centering
    \includegraphics[width=0.48\textwidth]{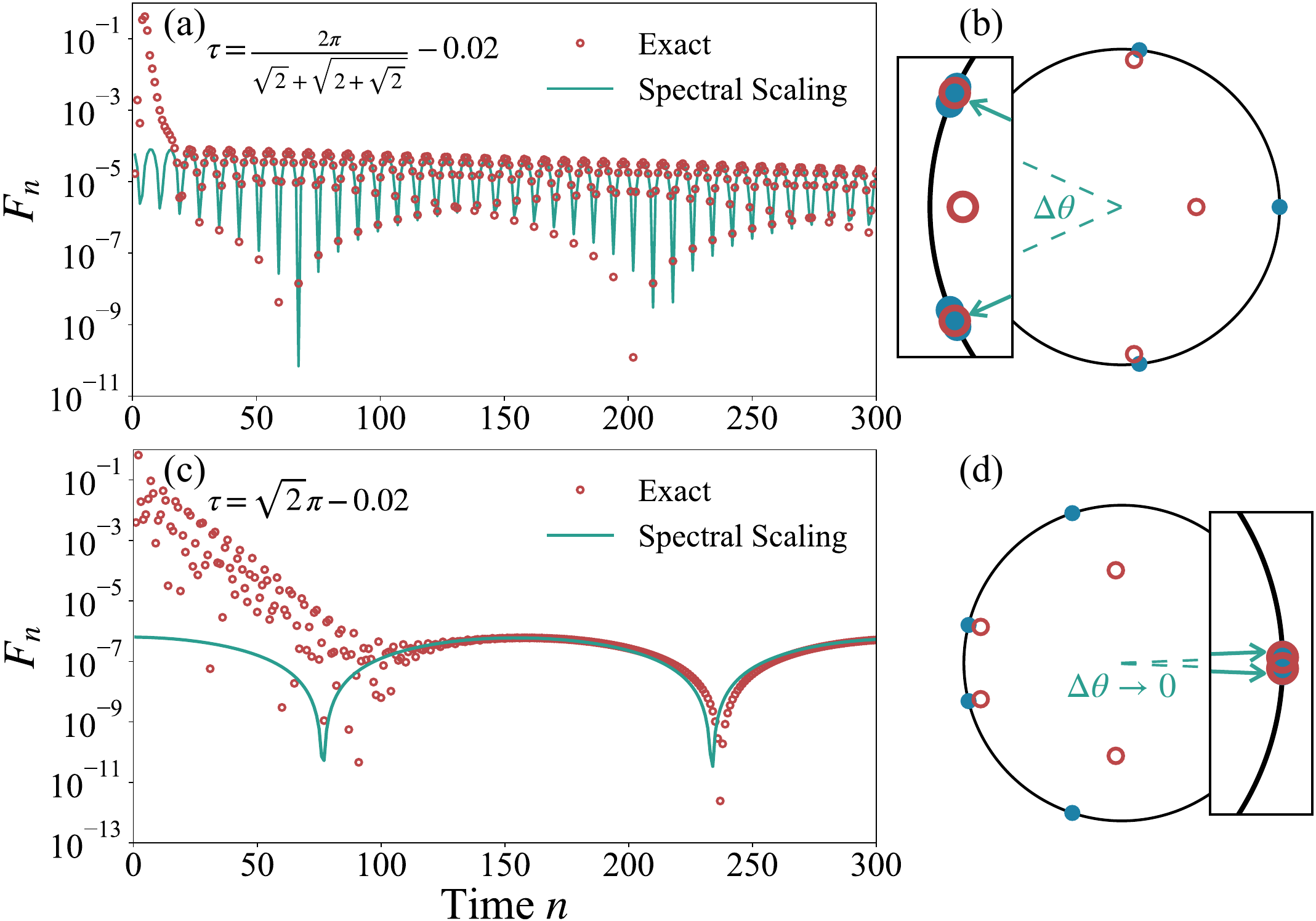}
%    \caption{First-detection probability $F_n$ versus measurement time $n$ for a seven-site tight-binding chain, illustrating two distinct oscillation regimes. (a)~Two-charge-pair resonance at $\tau = 2\pi/(\sqrt{2}+\sqrt{2+\sqrt{2}}) - 0.02$. Exact results (circles) computed from Eq.~\eqref{eq:phi_n} agree well with the spectral approximation of Eq.~\eqref{eq:Fn_spectral} (solid line) at large $n$. (b)~Corresponding unitary phase factor configuration (blue dots) and eigenvalues of $\mathcal{S}$ (red circles) within the unit disk. The two dominant eigenvalues of $\mathcal{S}$ approach the unit circle, and their phase difference $\Delta\theta$ sets the oscillation frequency. (c)~$F_n$ versus $n$ for three-charge coalescence at $\tau = \sqrt{2}\pi - 0.02$, where the oscillation period becomes extremely long, yielding an extended plateau in $F_n$. (d)~Corresponding unitary phase factor and eigenvalue distribution. The two dominant eigenvalues nearly coalesce, causing the oscillation frequency to vanish.}
% [QW: caption rephrased to avoid the ``charge'' terminology; wording now matches the section headers.]
    \caption{First-detection probability $F_n$ versus measurement time $n$ for a seven-site tight-binding chain, illustrating two distinct oscillation regimes. (a)~Merging of two pairs of phase factors at $\tau = 2\pi/(\sqrt{2}+\sqrt{2+\sqrt{2}}) - 0.02$. Exact results (circles) computed from Eq.~\eqref{eq:phi_n} agree well with the spectral approximation of Eq.~\eqref{eq:Fn_spectral} (solid line) at large $n$. (b)~Corresponding unitary phase factor configuration (blue dots) and eigenvalues of $\mathcal{S}$ (red circles) within the unit disk. The two dominant eigenvalues of $\mathcal{S}$ approach the unit circle, and their phase difference $\Delta\theta$ sets the oscillation frequency. (c)~$F_n$ versus $n$ for the merging of three phase factors at $\tau = \sqrt{2}\pi - 0.02$, where the oscillation period becomes extremely long, yielding an extended plateau in $F_n$. (d)~Corresponding distribution of unitary phase factors and eigenvalues of $\mathcal{S}$. The two dominant eigenvalues nearly coalesce, causing the oscillation frequency to vanish.}
    \label{fig:Fn_spectral}
\end{figure}

Bipartite symmetry allows for a second scenario if a zero-energy stationary state exists. This corresponds to bipartite systems with an odd number of energy levels. 
If an energy level is exactly $E=0$, the corresponding phase factor of the unitary operator $U(\tau)$ is clearly unity. For specific values of $\tau$, two additional phase factors, $e^{-iE\tau}$ and $e^{iE\tau}$, can be tuned to merge with the zero-energy phase factor and with each other, such that $e^{-iE\tau} = e^{iE\tau} = 1$. This condition can be easily found as the non-zero energy levels $E$ and $-E$ come in pairs,
as illustrated in Fig.~\ref{fig:Fn_spectral}(d).
Note that in Fig.~\ref{fig:Fn_spectral} the unitary phase factors $e^{-iE\tau}$ are shown as blue dots on the unit circle, not to be confused with the eigenvalues of ${\cal S}$ marked as red circles, which lie inside the unit disk.
Physically, this corresponds to the two dominant eigenvalues of $\mathcal{S}$ not only approaching the unit circle but also merging with each other,
producing an oscillation pattern qualitatively distinct from the finite-frequency fringes found above. 
A perturbative approach, using the deviation of the phase factors $e^{\pm iE\tau}$ from unity as the small parameter, then gives the long-time dynamics
\begin{equation}\label{eq:Fn_triple}
    F_n \sim 4 \left[ \lambda' (\widetilde{\Delta E \tau})^2 \right]^{2} e^{-\lambda' (\widetilde{\Delta E \tau})^2 n} \cos^2 \left( \frac{n}{2} \Delta \theta_{\text{slow}} + \beta \right),
\end{equation}
where $\lambda' = P_{(E)}/(P_{(E=0)}+2P_{(E)})^2$ and here $\Delta E=E-0$ is the energy difference measured from the zero-energy level. 
%In this case the zeroth-order term in Eq.~\eqref{eq:leading_theta} vanishes, and 
Here the oscillation frequency scales linearly with the detuning as
\begin{equation}\label{eq:theta_slow}
\Delta \theta_{\text{slow}} = -2\sigma \widetilde{\Delta E \tau},
\end{equation}
with an additional geometric factor $\sigma = \sqrt{P_{(E=0)}/(P_{(E=0)}+2P_{(E)})}$.
Since the detuning $\widetilde{\Delta E \tau}$ is small, so is this frequency, which creates a wide temporal window where the detection probability remains nearly constant, effectively ``freezing'' the decay process. Fig.~\ref{fig:Fn_spectral}(c) displays this plateau, with the spectral prediction (solid line) matching the exact dynamics (circles).

% In both cases the envelope decay scales as $(\Delta E\tau)^2$, while the frequency is governed by distinct energy scales, i.e.  the centroid splitting $\Delta \theta^{(0)}$ versus the detuning $\Delta E\tau$,
% thus envelope and fringe pattern can be tuned independently.

% The two regimes represent physically distinct forms of temporal interference. In the first, two well-separated pairs of nearly degenerate phase factors create two long-lived decay modes, namely two ``slits'' in time,
% whose fixed phase difference sets the observed fringes at a finite frequency, robust against detuning.
% In the second, the coalescence of three phase factors fuses these slits: the oscillation frequency itself vanishes linearly with detuning, converting interference fringes into a plateau where detection is nearly frozen.
% In both cases the envelope decay scales as $(\widetilde{\Delta E\tau})^2$, while the frequency is governed by distinct energy scales, i.e.  the splitting between the two resonant pairs versus the detuning $\widetilde{\Delta E\tau}$, thus envelope and fringe pattern can be tuned independently.
% [QW: rewrote the two-regime summary per EB comments -- recast the contrast as "frequency set by energy levels (Eq.7) vs by overlaps through sigma (Eq.11)"; dropped slit-reuse and "robust against detuning"]
The two regimes are physically distinct. When two pairs of phase factors merge, the two long-lived modes act as two ``slits'' in time, and their fixed phase difference sets the observed fringes: the frequency, Eq.~\eqref{eq:leading_theta}, depends on the energy levels alone, so the fringes directly probe the spectrum of $H$. When three phase factors of the unitary merge, this level contribution vanishes, and the much smaller frequency of Eq.~\eqref{eq:theta_slow} is now proportional to the detuning, with a prefactor set by the overlaps of the target state with the energy eigenstates; the fringes then stretch into the plateau of Fig.~\ref{fig:Fn_spectral}(c). In both regimes $F_n$ decays slowly, like $e^{-\lambda(\widetilde{\Delta E\tau})^2 n}$, with a lifetime that diverges as the transition is approached.
Explicit derivations of all coefficients are given in Appendix.

\section{Discussion}
In summary, we have uncovered a spectral origin of long-lived temporal interference in monitored quantum walks, emerging close to the topological transition $w\to w-2$. These transitions do not merely produce discrete jumps in the quantized mean recurrence time; they also reorganize the transient dynamics of the monitored system.
% Near the transition, the late-time dynamics is governed by two coherent decay modes.
% Remarkably, these modes have the same weights and locked phases, in analogy with a two-channel interference problem; their superposition produces high-visibility temporal fringes.
% Equivalently, this regime corresponds to the formation of a pair of long-lived quasi-dark states. The perturbative analysis developed in SM shows how small detunings from the ideal resonance move these modes inside the unit circle, thereby setting the slow decay of the interference envelope.
% Crucially, this mechanism is structurally protected by the bipartite symmetry of the underlying graph, whose mirror-symmetric spectrum guarantees the balanced weights and conjugate phases of the two dominant modes.
% Thus, bipartite symmetry provides a generic route to stable long-lived temporal interference in monitored quantum systems.
% [QW: compressed the five sentences above into three -- weights/phases now stated once and attributed to bipartite symmetry; dropped the SM-perturbation recap already given in the results section]
Near the transition, the late-time dynamics is governed by two coherent decay modes with equal weights and locked phases; as in a two-channel interference problem, their superposition produces high-visibility temporal fringes.
Equivalently, this regime corresponds to the formation of a pair of long-lived quasi-dark states.
Crucially, this mechanism is structurally protected by the bipartite symmetry of the underlying graph, whose mirror-symmetric spectrum guarantees the balanced weights and conjugate phases of the two dominant modes, providing a generic route to stable temporal interference in monitored quantum systems.
In particular, the fringe frequency distinguishes the two scenarios: in bipartite systems with a stationary $E=0$ we find slow oscillations, as in Fig.~\ref{fig:Fn_spectral}(c), whereas the merging of two pairs of phase factors yields faster oscillations, see Fig.~\ref{fig:Fn_spectral}(a).

Looking forward, observing these long-lived interference fringes offers a timely experimental opportunity. Current platforms, including trapped-ion systems \cite{Ryan2025}, photonic lattices \cite{Liu2023Feng,Liu2025Feng,Linda2026}, neutral atom arrays \cite{Deist2022,Graham2023}, and superconducting circuits \cite{Koh2023,Sabine2022,Wang2023,yin2024restart,Shuanger2026,Perfetto2026}, provide promising arenas to test our predictions. A critical next step is to investigate how realistic environmental noise and finite coherence times, ubiquitous in these implementations, modify or smear the interference patterns identified here. Extending these ideas to many-body systems \cite{Bourgain2014,Andreas2023,Purkayastha2024,Walter2025pra,Liu2024ManyBody,Yin2025_cat_states,Perfetto2026c,Kazuki2026,Perfetto2026b} presents another important challenge.

\section*{Acknowledgments}
The support of Israel Science Foundation's grant 2311/25 is acknowledged.

\section*{Author contributions}
Q.W., R.Y. and E.B. conceived the research. Q.W. and R.Y. did the analytical calculations and performed the numerical simulations. E.B. supervised the work on the project. All authors contributed to writing the manuscript.
Large Language Model (AI) tools were used for language polishing and editing. 

\bibliography{ref}

\appendix

\section{Charge theory}\label{app:charge}
The eigenvalues of the survival operator $\mathcal{S} = (\mathbb{1}-|x_T\rangle\langle x_T|)U(\tau)$ introduced in Eq.~\eqref{eq:phi_n} can be analytically determined through an electrostatic mapping \cite{Gruenbaum2013}. Applying the matrix determinant lemma to the characteristic polynomial of $\mathcal{S}$ yields \cite{Thiel2020D}
\begin{equation}
    0 = \text{det} \left[ \xi \mathbb{1} - \mathcal{S} \right] = \text{det} [ \xi \mathbb{1} - U(\tau) ] \langle x_T | [ \xi \mathbb{1} - U(\tau) ]^{-1} | x_T \rangle \xi.
\end{equation}
Evaluating the matrix element via spectral decomposition in the energy basis, we find that the non-trivial eigenvalues $\xi$ strictly inside the unit disk, which enter the spectral decomposition of the first-detection amplitude in Eq.~\eqref{eq:phi_n_decom}, are given by the roots of
\begin{equation}
    \sum_{k=1}^w \sum_{m=1}^{g_k} |\langle x_T|E_{k,m}\rangle|^2  \frac{1}{e^{-iE_k\tau}-\xi} = 0,
\end{equation}
where $\ket{E_{k,m}}$, $m=1,\dots,g_k$, are the degenerate eigenstates of energy $E_k$ in the notation of the main text, while $g_k$ denotes the degeneracy of the $k$-th energy level. Using the definition from the main text for the spectral weight of each distinct unitary phase factor, $P_k = \sum_{m=1}^{g_k} |\langle x_T|E_{k,m}\rangle|^2$, the eigenvalue equation simplifies to
\begin{equation}\label{eq:app_eigen}
    \sum_{k=1}^w \frac{P_k}{e^{-i E_k \tau}-\xi} = 0.
\end{equation}

This equation admits an intuitive electrostatic interpretation \cite{Gruenbaum2013,yin2019}. Each term $P_k/(e^{-i E_k \tau}-\xi)$ can be interpreted as the two-dimensional Coulomb-like force exerted by a positive fixed ``charge'' of magnitude $P_k$. The corresponding force field is given by
\begin{equation}\label{eq:force_field}
    \mathcal{F}(\xi) = \sum_{k=1}^w \frac{P_k}{e^{-i E_k \tau}-\xi},
\end{equation}
whose stationary points, determined by $\mathcal{F}(\xi)=0$, define the relevant eigenvalues of $\mathcal{S}$. Integration with respect to $\xi$ yields the corresponding Coulomb-like potential
\begin{equation}
    V(\xi) = \sum_{k=1}^w P_k \ln |e^{-i E_k\tau}-\xi|.
\end{equation}
Physically, this describes a two-dimensional cross-section of infinite parallel wires with uniform line charge densities $P_k$, threading perpendicularly through the unit plane at locations $e^{-i E_k \tau}$.
Equivalently, Eq.~\eqref{eq:force_field} shows that the sought-after eigenvalues of $\mathcal{S}$ inside the unit disk are the equilibrium points of this classical force field.

This mapping establishes a rigorous geometric correspondence between the energy spectrum and the charge configuration:
\begin{equation*}
    e^{-i E_k \tau} \leftrightarrow  \text{location of ``charge'' } P_k \text{ on the unit circle}.
\end{equation*}
The number of distinct charges corresponds exactly to the winding number $w$. By tuning the sampling time $\tau$, we rotate these charges along the unit circle. When the system approaches a topological resonance, distinct charges coalesce.
Two nearby charges always trap an equilibrium point of the force field between them, in analogy with the Laplace points of an electrostatic configuration \cite{Gruenbaum2013,yin2019}; as the charges merge, this equilibrium point is driven toward the unit circle.
Specifically, near a $w \to w-2$ transition, the simultaneous merging of two charge pairs sculpts a local force field that draws two equilibrium points ($\xi_1$ and $\xi_2$) extremely close to the unit boundary, thereby triggering the temporal interference discussed in the main text.
Building on this electrostatic picture, combined with perturbation theory in the small charge separation, we derive in Appendix~\ref{sec:sm_perturbation} the spectral forms of the first-detection probability quoted in the main text, Eqs.~\eqref{eq:Fn_spectral} and~\eqref{eq:Fn_triple}, together with the coefficients $\lambda$, $\eta$, $\sigma$.

% % [deleted: this appendix duplicates Eqs. (sm_coefficients) and (sm_beta) of the merged derivation below]
% \begingroup\color{gray}
% \section{Explicit expressions for the interference parameters}
% The spectral decomposition of the first-detection amplitude in Eq.~\eqref{eq:phi_n_decom} is governed by the complex coefficients $r_k e^{i\beta_k}$. These coefficients are analytically determined by the eigenvalues $\xi_k$. Evaluating these contributions yields the exact expressions for the spectral weights:
% \begin{equation}\label{eq:app_coefficients}
%     r_k e^{i\beta_k} = -e^{-i \zeta}\frac{\prod_{j = 1}^{w-1} ( 1 - \xi_j^* \xi _k )}{(\xi_k)^2 \prod_{j \neq k} ( \xi _j - \xi _k )}, 
% \end{equation}
% where $\zeta = \sum_{\ell=1}^w E_\ell \tau$ is the total phase accumulated by the unitary dynamics.

% The global phase shift $\beta$ appearing in the interference pattern [Eq.~\eqref{eq:interference1}] is explicitly given by:
% \begin{equation}\label{eq:app_beta}
%     \begin{split}
%     \beta = \frac{1}{2}\mathrm{Arg}\left[\frac{\prod_{j = 1}^{w-1} ( 1 - \xi_j^* \xi _1 )}{(\xi_1)^2 \prod_{j \neq 1} ( \xi _j - \xi _1 )}\right] &-\\
% \frac{1}{2}\mathrm{Arg}&\left[\frac{\prod_{j = 1}^{w-1} ( 1 - \xi_j^* \xi _2 )}{(\xi_2)^2 \prod_{j \neq 2} ( \xi _j - \xi _2 )}\right].
%     \end{split}
% \end{equation}
% Recall that $\xi_1$ and $\xi_2$ are the two eigenvalues of $\mathcal{S}$ approaching the unit circle.
% \endgroup

\section{Monotonic decay at $w \to w-1$ transitions}\label{app:monotonic}
The $w \to w-1$ transition involves only a single eigenvalue of $\mathcal{S}$ approaching the unit circle, as illustrated in Fig.~\ref{fig:Fn_4hypercube}(c). Suppose $\xi_1$ is this sole dominant eigenvalue with $|\xi_1| \lesssim 1$, while all other eigenvalues satisfy $|\xi_{k>1}| \ll |\xi_1|$. 

In the long-time limit ($n \gg 1$), the transient contributions from the faster-decaying modes in the spectral decomposition [Eq.~\eqref{eq:phi_n_decom}] become exponentially suppressed. The first-detection amplitude is then entirely governed by this single slow mode:
\begin{equation}
    \phi(n, x_T) \simeq r_1 e^{i\beta_1} \xi_1^n.
\end{equation}
Consequently, the first-detection probability loses all interference cross-terms, reducing to a purely monotonic exponential decay \cite{yin2024restart}:
\begin{equation}
    F_n = |\phi(n, x_T)|^2 \sim \big(1 - |\xi_1|^2 \big)^2 \, |\xi_1|^{2n}.
\end{equation}
Without a partner mode of comparable magnitude to generate a relative phase, $F_n$ carries no oscillations and decays as a single exponential. This single-mode asymptotic behavior analytically explains the slow monotonic decay observed in Fig.~\ref{fig:fig_protocol1}(a) (blue triangles) and aligns with the isolated quasi-dark state configuration shown in Fig.~\ref{fig:Fn_4hypercube}(c).

% \textit{Appendix D: Continuous-$n$ plotting of the theoretical curves}---
% The exact data (circles) in Figs.~\ref{fig:Fn_4hypercube} and~\ref{fig:Fn_spectral} are computed from Eq.~\eqref{eq:phi_n} at the integer steps $n=1,2,\ldots$, where $F_n$ is physically defined. The theoretical curves (solid lines), however, are plotted as functions of a \emph{continuous} variable $n$. Since our asymptotic formulas carry a factor $\cos^2[\tfrac{n}{2}\Delta\theta-\beta]$, the continuous curve necessarily touches zero at its nodes, whereas the exact data, sampled only at integer $n$, generically fall between these nodes and therefore essentially never vanish. This explains why, at the minima of the interference fringes, the solid line dips to zero while the circles remain slightly above it. (Note also that the theory in the two figures is different: Fig.~\ref{fig:Fn_4hypercube} uses Eq.~\eqref{eq:interference1} with inputs taken directly from the eigenvalues of $\mathcal{S}$, while Fig.~\ref{fig:Fn_spectral} uses the spectral approximations Eqs.~\eqref{eq:Fn_spectral} and~\eqref{eq:Fn_triple}, whose parameters are expressed through the spectrum of the Hamiltonian.)

\section{The hypercube model}\label{app:hypercube}
The $d$-dimensional hypercube, which is a bipartite graph, provides a natural representation of a quantum register consisting of $d$ qubits and serves as a fundamental model in the study of quantum hitting times and quantum computing \cite{kempe2005, Krovi2006a, Krovi2006, Thiel2020D, Howard2019, Goto2024}. 
The vertices of the graph are the computational basis states 
$|z\rangle = \bigotimes_{j=1}^d |z_j\rangle$ with $z_j \in \{0,1\}$, 
and nearest-neighbor hopping corresponds to single spin flips.
The tight-binding Hamiltonian is therefore that of $d$ noninteracting 
spins in a transverse field,
\begin{equation}\label{eq:hypercube_ham}
    H = -\gamma \sum_{j=1}^{d} \hat{\sigma}_x^{(j)},
\end{equation}
where we set the hopping amplitude $\gamma=1$. 
The energy levels are $E_k = -(d-2k)$ for $k=0,\dots,d$, with degeneracy
$g_k = \binom{d}{k}$. For the recurrence problem initialized at a single
vertex, $|x_T\rangle = |0\rangle^{\otimes d}$, the spectral weights are
binomial, $P_k = \binom{d}{k}\,2^{-d}$.
The spectrum and weights satisfy $E_k = -E_{d-k}$ and $P_k = P_{d-k}$,
realizing the bipartite symmetry that protects the interference
visibility discussed in the main text.
The specific values of $P_k$ for the $d=4$ hypercube used in the main text are given in  Appendix \ref{sec:sm_parameters}.

\section{The tight-binding chain model}\label{app:chain}
We consider a one-dimensional tight-binding chain of $L$ sites with open boundary conditions. The Hamiltonian is given by
\begin{equation}\label{eq:chain_ham}
H = -\gamma \sum_{j=1}^{L-1} \big(|j\rangle\langle j+1| + \text{h.c.}\big),
\end{equation}
where $\gamma$ is the hopping amplitude (set to $\gamma=1$). The system possesses a discrete spectrum $E_k = -2\gamma \cos(\frac{k\pi}{L+1})$ for $k=1, \dots, L$.
This spectrum exhibits the particle-hole symmetry $E_k = -E_{L+1-k}$ inherent to the bipartite structure discussed in the main text.
In contrast to the hypercube, the chain has no energy degeneracy: each level is nondegenerate, $g_k=1$.
Focusing on the chain model, for a walker initialized at an edge site the spectral weights $P_k=|\langle 1 | E_k \rangle|^2 = \frac{2}{L+1} \sin^2(\frac{k\pi}{L+1})$ are non-vanishing for all $k$. 
Since all $L$ energy levels are distinct and contribute to the first-detection amplitude of Eq.~\eqref{eq:phi_n}, the generic winding number corresponds to the system dimension, $w=L$.
For odd $L$ a zero-energy level exists, and the chain realizes the three-charge scenario of Fig.~\ref{fig:Fn_spectral}(c,d), in which this level merges with a pair $\pm E$. The specific weights $P_{(E=0)}$ and $P_{(E)}$ entering the coefficients $\lambda'$ and $\sigma$ of Eq.~\eqref{eq:Fn_triple}, together with the perturbation theory for three coalescing charges, are given in  Appendices \ref{sec:sm_perturbation} and \ref{sec:sm_parameters}.

\section{Generating function of the first-detection amplitude}\label{sec:sm_gf}

%\subsection{Solving the quantum renewal equation}

Our starting point is the first-detection amplitude, Eq.~\eqref{eq:phi_n},
\begin{equation}\label{eq:sm_phi_def}
\phi(n,x_T)=\bra{x_T}U(\tau)\,\mathcal{S}^{\,n-1}\ket{x_T}.
\end{equation}
The amplitudes obey the quantum renewal equation \cite{Gruenbaum2013,Friedman2017a}
%
% [QW: per E.B. -- renamed the intermediate-attempt summation index m -> s here and in the convolution below, to free the letter m for the level degeneracy in |E_{k,m}>; also put a full stop after the equation and rearranged the following sentence to state the measurement-free return amplitude directly.]
%\begin{equation}\label{eq:sm_QRE}
%  \begin{split}
%    \phi(n,x_T)=\bra{x_T} U(n \tau)\ket{x_T}
%    -\sum_{m=1}^{n-1}\bra{x_T} U[(n-m) \tau]\ket{x_T}\, \phi(m,x_T),
%  \end{split}
%\end{equation}
%
\begin{equation}\label{eq:sm_QRE}
  \begin{split}
    \phi(n,x_T)=\bra{x_T} &U(n \tau)\ket{x_T} - \\ &\sum_{s=1}^{n-1}\bra{x_T} U[(n-s) \tau]\ket{x_T}\, \phi(s,x_T).
  \end{split}
\end{equation}
%
%which expresses the measurement-free return amplitude at time $n\tau$ as the sum over mutually exclusive first-detection events: either the state is detected for the first time at the $n$-th attempt, or it was first detected at some earlier attempt $m<n$ and then freely propagated back to $\ket{x_T}$ during the remaining time $(n-m)\tau$.
Rearranging the terms, the measurement-free return amplitude $\bra{x_T}U(n\tau)\ket{x_T}$ at time $n\tau$ is the sum over mutually exclusive first-detection events: either the state is detected for the first time at the $n$-th attempt, or it was first detected at some earlier attempt $s<n$ and then freely propagated back to $\ket{x_T}$ during the remaining time $(n-s)\tau$.

To solve Eq.~\eqref{eq:sm_QRE} we introduce the generating function
\begin{equation}\label{eq:sm_gf_def}
\hat{\phi}(z) = \sum_{n=1}^\infty z^n \phi(n,x_T).
\end{equation}
Multiplying Eq.~\eqref{eq:sm_QRE} by $z^n$ and summing over $n$, the first term gives $\bra{x_T}\hat{U}(z)\ket{x_T}$ with $\hat{U}(z) = \sum_{n=1}^\infty z^n U(n\tau)$, while the second term is a discrete convolution,
%
% [QW: per E.B. -- summation index m -> s here as well, for consistency with Eq.(S2).]
%\begin{equation}
%\begin{aligned}
%\sum_{n=1}^{\infty} z^n \sum_{m=1}^{n-1} \bra{x_T} U[(n-m) \tau]\ket{x_T} \phi(m,x_T)= \sum_{m=1}^{\infty} \phi(m,x_T)\, z^m \sum_{n'=1}^{\infty} z^{n'} \bra{x_T} U(n' \tau)\ket{x_T}=  \hat{\phi}(z)\, \bra{x_T}\hat{U}(z)\ket{x_T},
%\end{aligned}
%\end{equation}
%
% \begin{widetext}
\begin{equation}
    \begin{split}
&\sum_{n=1}^{\infty} z^n \sum_{s=1}^{n-1} \bra{x_T} U[(n-s) \tau]\ket{x_T} \phi(s,x_T)\\ =& \sum_{s=1}^{\infty} \phi(s,x_T)\, z^s \sum_{n'=1}^{\infty} z^{n'} \bra{x_T} U(n' \tau)\ket{x_T}\\ = & \hat{\phi}(z)\, \bra{x_T}\hat{U}(z)\ket{x_T},
\end{split}
\end{equation}
% \end{widetext}
%
with $n'=n-s$. Solving for $\hat{\phi}(z)$ yields \cite{Gruenbaum2013,Friedman2017a}
\begin{equation}\label{eq:sm_gf_solution}
\hat{\phi}(z) =
\frac{\bra{x_T} \hat{U}(z) \ket{x_T}}
{1 + \bra{x_T} \hat{U}(z) \ket{x_T}}.
\end{equation}

%\subsection{Spectral representation and pole--zero structure}

Writing $U(\tau)\ket{E_{k,m}}=e^{-iE_k\tau}\ket{E_{k,m}}$, where $k=1,\dots,w$ labels the distinct phase factors $e^{-iE_k\tau}$ and $m$ their possible degeneracy, and summing the geometric series in Eq.~\eqref{eq:sm_gf_def}, we obtain
\begin{equation}\label{eq:sm_U_matrix_element}
  \bra{x_T} \hat{U}(z) \ket{x_T} =
  \sum_{k=1}^w P_k \, \frac{z}{e^{i E_k \tau} - z},
\end{equation}
with the spectral weights $P_k=\sum_{m=1}^{g_k} |\langle x_T|E_{k,m}\rangle|^2$ defined in Appendix~\ref{app:charge}. For simplicity we assume $P_k\neq0$; our main conclusions are unchanged otherwise.
Inserting Eq.~\eqref{eq:sm_U_matrix_element} into Eq.~\eqref{eq:sm_gf_solution} and using the normalization $\sum_k P_k = 1$, the generating function becomes a ratio of polynomials, $\hat{\phi}(z)={\cal N}(z)/{\cal D}(z)$, with
\begin{equation}\label{eq:sm_ND}
\begin{aligned}
{\cal N}(z) &= z \sum_{k=1} ^w P_k \prod_{j \neq k} ^w \big( e^{ i E_j \tau}- z \big), \\
{\cal D}(z) &= \sum_{k=1} ^w P_k\, e^{ i E_k \tau } \prod_{j\neq k} ^w \big( e^{ i E_j \tau} - z \big).
\end{aligned}
\end{equation}
The numerator is a polynomial of degree $w$ with a trivial zero at $z=0$, hence
\begin{equation}\label{eq:sm_N_factorized}
{\cal N}(z) = z \prod_{i=1}^{w-1} (z_i - z),
\end{equation}
and the $w-1$ non-trivial zeros $z_i$, which all lie strictly inside the unit disk \cite{Gruenbaum2013}, satisfy
\begin{equation}\label{eq:sm_zero_condition}
    \sum_{k=1}^w \frac{P_k}{e^{i E_k \tau}-z_i} = 0 .
\end{equation}
%
%Equation~\eqref{eq:sm_zero_condition} is precisely the complex conjugate of the eigenvalue condition for the survival operator, End Matter Eq.~(14), since the weights $P_k$ are real. Therefore
% [QW: per E.B. -- sentence was unclear; state directly that Eq.(S9) is the complex conjugate of End Matter Eq.(14).]
Eq.~\eqref{eq:sm_zero_condition} is the complex conjugate of Eq.~\eqref{eq:app_eigen}, the eigenvalue condition for the survival operator $\mathcal{S}$. Therefore
\begin{equation}\label{eq:sm_conjugation}
z_i = \xi_i^*, \qquad i=1,\dots,w-1,
\end{equation}
i.e., the non-trivial zeros of the generating function are the complex conjugates of the non-trivial eigenvalues of $\mathcal{S}$. This is the dictionary between the eigenvalue language of the main text and the zeros language of Refs.~\cite{Gruenbaum2013,yin2019}.

The denominator is related to the numerator by the reflection identity ${\cal D}(z) = (-1)^{w-1} e^{i\zeta} z^w \left[{\cal N}(1/z^*)\right]^*$, where $\zeta = \sum_{\ell=1}^w E_\ell\tau$ is the total phase accumulated by the distinct levels, as can be verified directly from the definitions in Eq.~\eqref{eq:sm_ND}. Combining this identity with Eqs.~\eqref{eq:sm_N_factorized} and \eqref{eq:sm_conjugation}, the generating function takes the factorized form \cite{Gruenbaum2013,yin2019}
\begin{equation}\label{eq:sm_gf_factorized}
\hat{\phi}(z)=
e^{- i \zeta} \, \frac{  z \prod_{i=1}^{w-1} (z - \xi_i^*) }
{\left( \prod_{i=1}^{w-1} \xi_i \right) \prod_{i=1}^{w-1} \big(z - 1/\xi_i \big)}.
\end{equation}
The poles of $\hat{\phi}(z)$ are located at $1/\xi_i$, outside the unit disk, and they control the large-$n$ decay of the amplitude: the closer an eigenvalue $\xi_i$ is to the unit circle, the closer the corresponding pole is to it from the outside, and the slower the associated decay.

% \section{Spectral decomposition of the amplitude}
\section{Spectral decomposition of the amplitude \texorpdfstring{[Eq.~\eqref{eq:phi_n_decom}]}{}}\label{sec:sm_decomposition}

The first-detection amplitude $\phi(n,x_T)$ is recovered from the generating function by the Cauchy integral
\begin{equation}\label{eq:sm_cauchy}
\phi(n,x_T) = { 1 \over 2 \pi i} \oint_{|z|=1} \hat{\phi}(z)\, z^{- (n+1)} {\rm d} z .
\end{equation}
%
% [QW: polished/merged -- removed the repeated "substituting Eq.(S11)" steps and the standalone residue equation, folding its value into the coefficients (S16) below.]
%\newc{Substituting Eq.~\eqref{eq:sm_gf_factorized} into Eq.~\eqref{eq:sm_cauchy} and using the residue theorem yield}
The integrand has poles at $z=1/\xi_k$, outside the unit circle [Eq.~\eqref{eq:sm_gf_factorized}]. Deforming the contour outward, the integral is given by minus the sum of the residues at these poles,
\begin{equation}\label{eq:sm_residue_sum}
\phi(n,x_T) = -\sum_{k=1}^{w-1}\underset{z=1/\xi_k}{\mathrm{Res}}\left[\hat{\phi}(z)\, z^{-(n+1)}\right].
\end{equation}
%
%Using Eq.~\eqref{eq:sm_gf_factorized} , the residue at $z=1/\xi_k$ reads
%\begin{equation}
%\underset{z=1/\xi_k}{\mathrm{Res}}
%= e^{-i\zeta}\,
%\frac{\prod_{j = 1}^{w-1} ( 1 - \xi_j^* \xi _k )}{(\xi_k)^2 \prod_{j \neq k} ( \xi _j - \xi _k )}\;
%\xi_k^{\,n}.
%\end{equation}
%Substituting into Eq.~\eqref{eq:sm_residue_sum}, we arrive at the spectral decomposition quoted in Eq.~(3) of the Letter,
Evaluating each residue with Eq.~\eqref{eq:sm_gf_factorized} gives the spectral decomposition quoted in Eq.~\eqref{eq:phi_n_decom},
\begin{equation}\label{eq:sm_decomposition_result}
\phi(n,x_T) = \sum_{k=1}^{w-1} r_k e^{i\beta_k}\,\xi_k^{\,n},
\end{equation}
with the coefficients
\begin{equation}\label{eq:sm_coefficients}
    r_k e^{i\beta_k} = -e^{-i \zeta}\frac{\prod_{j = 1}^{w-1} ( 1 - \xi_j^* \xi _k )}{(\xi_k)^2 \prod_{j \neq k} ( \xi _j - \xi _k )}.
\end{equation}
In the notation of the main text, the sum in Eq.~\eqref{eq:phi_n_decom} also includes the trivial eigenvalue $\xi=0$, which does not contribute to the sum.

% \newc{The residue sum \eqref{eq:sm_residue_sum} was derived for $n\geq2$, where the integrand decays as $z^{-n}$ and the outer contour drops out. For $n=1$ the integrand falls off only as $z^{-1}$, so the arc at infinity contributes an extra term and the pole sum \eqref{eq:sm_decomposition_result} no longer holds; in this case the amplitude is given directly by Eq.~\eqref{eq:sm_phi_def}, $\phi(1,x_T)=\bra{x_T}U(\tau)\ket{x_T}$. The two-mode analysis below concerns the large-$n$ regime, so this single initial value is immaterial.}

% \section{Two-mode asymptotics: derivation of Eq.~(5)}
\section{Two-mode interference asymptotics \texorpdfstring{[Eq.~\eqref{eq:interference1}]}{}}\label{sec:sm_two_mode}

We now consider the situation of interest in the main text: a pair of dominant eigenvalues,
\begin{equation}
\xi_1 = |\xi| e^{i\theta_1},\qquad \xi_2 = |\xi| e^{i\theta_2},
\end{equation}
with $|\xi|$ close to unity, while all other (subdominant) eigenvalues remain well inside the unit disk, $|\xi_{j\geq3}| < |\xi|$. For large $n$ the subdominant contributions to Eq.~\eqref{eq:sm_decomposition_result} are exponentially suppressed, and the amplitude reduces to the two-mode form of Eq.~\eqref{eq:two_mode_phi}. The remaining task is to evaluate the two coefficients $r_{1,2}e^{i\beta_{1,2}}$ in the regime $|\xi|\to1$.

%Isolating in Eq.~\eqref{eq:sm_coefficients} the factor associated with the second dominant eigenvalue $\xi_2$ from those of the subdominant modes, we write for $k=1$
In Eq.~\eqref{eq:sm_coefficients}, separating the factor associated with the second dominant eigenvalue $\xi_2$ from those of the subdominant modes, we write for $k=1$
\begin{equation}\label{eq:sm_r1_split}
    r_1 e^{i\beta_1} = -e^{-i\zeta}\,(1-|\xi|^2)\;
    \frac{1-\xi_2^*\xi_1}{\xi_2-\xi_1}\;
    \frac{1}{\xi_1^2}\;
    \prod_{j\geq 3}\frac{1-\xi_j^*\xi_1}{\xi_j-\xi_1},
\end{equation}
and similarly for $k=2$ with $1\leftrightarrow 2$. 

It is convenient to define, for a phase point $e^{i\theta}$ on the unit circle,
\begin{equation}\label{eq:sm_g_def}
g(\theta) =
\prod_{j \geq 3}
\frac{1 - \xi^{*}_j e^{i\theta}}{\xi_j - e^{i\theta}},
\end{equation}
which collects the influence of all subdominant eigenvalues. The key property of $g(\theta)$ is that it is a pure phase,
$|g(\theta)|=1$.

%In the limit $|\xi|\to1$ we may replace $\xi_{1,2}\to e^{i\theta_{1,2}}$ in all factors of Eq.~\eqref{eq:sm_r1_split} except the explicit prefactor $(1-|\xi|^2)$. The partner factor then simplifies to a pure phase as well,
We then use the limit $|\xi|\to1$, replacing $\xi_{1,2}\to e^{i\theta_{1,2}}$ in all factors of Eq.~\eqref{eq:sm_r1_split} except the explicit prefactor $(1-|\xi|^2)$. The factor coupling the two dominant modes then reduces to a pure phase as well,
\begin{equation}
\frac{1-\xi_2^*\xi_1}{\xi_2-\xi_1}\;\longrightarrow\;
\frac{1-e^{i(\theta_1-\theta_2)}}{e^{i\theta_2}-e^{i\theta_1}} = e^{-i\theta_2},
\end{equation}
so that we have
\begin{equation}\label{eq:sm_r_final}
\begin{aligned}
    r_1 e^{i\beta_1} &\simeq -e^{-i\zeta}(1-|\xi|^2)\, e^{-i(2\theta_1+\theta_2)}\, g(\theta_1),\\
    r_2 e^{i\beta_2} &\simeq -e^{-i\zeta}(1-|\xi|^2)\, e^{-i(2\theta_2+\theta_1)}\, g(\theta_2).
\end{aligned}
\end{equation}
Since $|g(\theta)|=1$, we conclude that
\begin{equation}\label{eq:sm_equal_weights}
r_1 = r_2 = 1-|\xi|^2 .
\end{equation}
This proves the statement made in the main text below Eq.~\eqref{eq:two_mode_phi}: although the condition $r_1=r_2$ appears \textit{a priori} unlikely, it is satisfied automatically whenever the pair of eigenvalues approaches the unit circle, i.e., close to the $w\to w-2$ topological transitions. The interference is therefore maximal precisely in the regime where it is long-lived. 
%In Sec.~\ref{sec:sm_bipartite} we show that for bipartite systems the equality $r_1=r_2$ is in fact exact for arbitrary $\tau$, and not only in the limit $|\xi|\to 1$.

%Inserting Eqs.~\eqref{eq:sm_r_final} into the two-mode sum,
Inserting Eqs.~\eqref{eq:sm_r_final} and \eqref{eq:sm_equal_weights} into the two-mode sum,
$\phi(n,x_T)\simeq r_1e^{i\beta_1}\xi_1^n + r_2e^{i\beta_2}\xi_2^n$, and computing $F_n=|\phi(n,x_T)|^2$, the global phase $-e^{-i\zeta}$ drops out and the cross term produces the interference fringes:
\begin{equation}\label{eq:sm_interference_final}
F_n \sim 4(1-|\xi|^2)^2\, |\xi|^{2n} \cos^2\left[\frac{n}{2}(\theta_1-\theta_2) + \beta\right],
\end{equation}
which is Eq.~\eqref{eq:interference1}, with the phase
\begin{equation}\label{eq:sm_beta}
\begin{split}
\beta = \frac{1}{2}\mathrm{Arg}\left[\frac{\prod_{j = 1}^{w-1} ( 1 - \xi_j^* \xi _1 )}{(\xi_1)^2 \prod_{j \neq 1} ( \xi _j - \xi _1 )}\right] &-\\
\frac{1}{2}\mathrm{Arg} &\left[\frac{\prod_{j = 1}^{w-1} ( 1 - \xi_j^* \xi _2 )}{(\xi_2)^2 \prod_{j \neq 2} ( \xi _j - \xi _2 )}\right].
\end{split}
\end{equation}
As noted in the main text, the phase $\beta$ does not admit a simple interpretation, in contrast to $|\xi|$ and the phase difference $\theta_1-\theta_2$, which together set the slow envelope decay of $F_n$ and the oscillation period.

\section{Perturbation theory near the $w\to w-2$ transitions}\label{sec:sm_perturbation}

So far the interference formula, Eq.~\eqref{eq:sm_interference_final}, is expressed through the two dominant eigenvalues of $\mathcal{S}$. We now relate $|\xi|$ and $\Delta\theta$ to the underlying energy spectrum and the overlaps $P_k=\sum_{m=1}^{g_k}|\langle x_T|E_{k,m}\rangle|^2$ through the electrostatic mapping of Appendix~\ref{app:charge}, in which the eigenvalues of $\mathcal{S}$ are the stationary points of the force field [Eq.~\eqref{eq:force_field}] generated by charges $P_k$ at the positions $e^{-iE_k\tau}$ on the unit circle.
The magnitude $|\xi|$ follows from perturbation theory in this framework~\cite{yin2019}, the small parameter being the detuning $\widetilde{\Delta E\,\tau}$ from resonance (defined below). The goal now is to find the phases $\theta_1$ and $\theta_2$, and hence the oscillation frequency of $F_n$, namely $\theta_1-\theta_2$, entering Eqs.~\eqref{eq:Fn_spectral}--\eqref{eq:theta_slow}.

\subsection{Merging of two pairs of phase factors \texorpdfstring{[Eqs.~\eqref{eq:leading_theta}--\eqref{eq:Fn_freq}]}{}}\label{sec:sm_case1}

Consider first the resonance scenario of the main text in which two pairs of phase factors merge independently,
$e^{-iE_1\tau}\simeq e^{-iE_2\tau}$ and $e^{-iE_3\tau}\simeq e^{-iE_4\tau}$,
while the two pairs, and all remaining phase factors, stay well separated from each other. We first treat the general case, in which the four levels and their weights are unrelated, and only then specialize to bipartite systems, for which the parameters of the two pairs unify into the form used in the main text.

\textit{General case.---}Each merging pair of charges traps one dominant eigenvalue near the unit circle. The perturbative result for the magnitudes is \cite{yin2019, yin2024restart}
\begin{equation}\label{eq:sm_moduli}
|\xi_1|^2 \simeq 1-\lambda_1 (\widetilde{\Delta E_1 \tau})^2,
\qquad
|\xi_2|^2 \simeq 1-\lambda_2 (\widetilde{\Delta E_2 \tau})^2,
\end{equation}
where $\Delta E_1 = E_1-E_2$ and $\Delta E_2 = E_4-E_3$ are the intra-pair splittings. Throughout, a tilde denotes reduction modulo $2\pi$ to $(-\pi,\pi]$, that is, $\widetilde{\Delta E\,\tau}\equiv\Delta E\,\tau\ (\mathrm{mod}\ 2\pi)$; this reduced phase measures the detuning of the sampling time from exact merging, and is the small parameter of the expansion. The decay coefficients read
\begin{equation}\label{eq:sm_lambdas}
\lambda_1 = \frac{P_1 P_2}{(P_1+P_2)^3},
\qquad
\lambda_2 = \frac{P_3 P_4}{(P_3+P_4)^3}.
\end{equation}
%
%Pronounced interference requires the equal-decay condition $|\xi_1|=|\xi_2|$, equivalently $\lambda_1(\widetilde{\Delta E_1\tau})^2 = \lambda_2(\widetilde{\Delta E_2\tau})^2$ by Eq.~\eqref{eq:sm_moduli}, so that the two dominant eigenvalues approach the unit circle together.
The sought-after long-lived interference pattern requires the equal-decay condition $|\xi_1|=|\xi_2|$, equivalently $\lambda_1(\widetilde{\Delta E_1\tau})^2 = \lambda_2(\widetilde{\Delta E_2\tau})^2$ by Eq.~\eqref{eq:sm_moduli}, so that the two dominant eigenvalues approach the unit circle together. As mentioned in the text, these conditions are naturally satisfied by bipartite graphs.

The phases are obtained by treating each pair in isolation. For two charges alone, located at $a_1 = e^{-iE_1\tau}$ and $a_2 = e^{-iE_2\tau}$ with strengths $P_1$ and $P_2$, the force balance $P_1/(a_1-\xi)+P_2/(a_2-\xi)=0$ is solved exactly by
\begin{equation}\label{eq:sm_two_charge}
\xi_1 \simeq \frac{P_1\, a_2 + P_2\, a_1}{P_1+P_2},
\end{equation}
i.e., the stationary point lies on the segment between the two charges, closer to the weaker one. Writing $E_1 = E_{12}+\Delta E_1/2$ and $E_2 = E_{12}-\Delta E_1/2$, with the centroid $E_{12}=(E_1+E_2)/2$, Eq.~\eqref{eq:sm_two_charge} can be rewritten as
\begin{equation}
\begin{aligned}
\xi_1 \simeq e^{-iE_{12}\tau}\left[\cos\frac{\widetilde{\Delta E_1\,\tau}}{2} + 2i\eta_1 \sin\frac{\widetilde{\Delta E_1\,\tau}}{2}\right] = e^{-iE_{12}\tau}\, R\, e^{i\Omega},
\end{aligned}
\end{equation}
where $\eta_1 = (P_1-P_2)/[2(P_1+P_2)]$, $R = \sqrt{\cos^2\frac{\widetilde{\Delta E_1\tau}}{2}+4\eta_1^2\sin^2\frac{\widetilde{\Delta E_1\tau}}{2}}$, and $\Omega = \arctan\left(2\eta_1\tan\frac{\widetilde{\Delta E_1\tau}}{2}\right)$. Hence the phase of the dominant eigenvalue is
\begin{equation}\label{eq:sm_theta1}
  \begin{split}
\theta_1 = \text{Arg}[\xi_1] = &-E_{12}\,\tau + \arctan\left(2\eta_1\tan\frac{\widetilde{\Delta E_1\,\tau}}{2}\right)  \\\simeq &-E_{12}\,\tau + \eta_1\, \widetilde{\Delta E_1\,\tau} ,
  \end{split}
\end{equation}
to first order in the detuning: the eigenvalue inherits the centroid phase of the merging pair, and the weight imbalance pulls it toward the weaker charge, in proportion to the detuning. %We use the isolated-pair solution \eqref{eq:sm_two_charge} only for the phase; the magnitude, which is more sensitive to the distant charges, is given by Eq.~\eqref{eq:sm_moduli}.
% [QW: per E.B. -- phrased as "use Eq.(S27) to derive Eq.(S29)"; also recall the separate fringe phase beta.]
We thus use the isolated-pair solution \eqref{eq:sm_two_charge} to derive the phase of $\xi_1$, Eq.~\eqref{eq:sm_theta1}, while the magnitude, which is more sensitive to the distant charges, is taken from Eq.~\eqref{eq:sm_moduli}.

Applying the same formula to the second pair, with centroid $E_{34}=(E_3+E_4)/2$ and splitting $\Delta E_2=E_4-E_3$, gives $\theta_2 \simeq -E_{34}\tau - \eta_2\,\widetilde{\Delta E_2\,\tau}$, with $\eta_2 = (P_3-P_4)/[2(P_3+P_4)]$. In the general case the phase difference is therefore
\begin{equation}\label{eq:sm_Delta_theta_general}
\Delta\theta \simeq
-(E_{12}-E_{34})\tau
+ \eta_1\,\widetilde{\Delta E_1\,\tau} + \eta_2\,\widetilde{\Delta E_2\,\tau} ,
\end{equation}
which involves the two splittings and the two weight-imbalance factors separately.

% [QW: per E.B. -- "Reduction" was unclear to the reader; renamed the paragraph and added a lead-in sentence.]
%\textit{Reduction by bipartite symmetry.---}For a bipartite system the two merging pairs are mutually conjugate, $E_3=-E_1$ and $E_4=-E_2$, and the spectral pairing derived in Sec.~\ref{sec:sm_bipartite} gives $P_3=P_1$ and $P_4=P_2$, as stated in the Letter.
\textit{Bipartite symmetry} of the underlying Hamiltonian $H$ further simplifies these expressions. For a bipartite system the two merging pairs are mutually conjugate, $E_3=-E_1$ and $E_4=-E_2$, and the spectral pairing derived in Appendix~\ref{sec:sm_bipartite} gives $P_3=P_1$ and $P_4=P_2$, as stated in the main text. The parameters of the two pairs then coincide,
\begin{equation}
\Delta E_1 = \Delta E_2 =: \Delta E,
\quad
\lambda_1=\lambda_2 =: \lambda,
\quad
\eta_1 = \eta_2 =: \eta,
\end{equation}
with $\lambda = P_1P_2/(P_1+P_2)^3$ and $\eta = (P_1-P_2)/[2(P_1+P_2)]$ as defined in the main text. Two consequences follow. First, by Eq.~\eqref{eq:sm_moduli} the equal-decay condition is satisfied automatically,
\begin{equation}\label{eq:sm_equal_moduli}
|\xi_1| = |\xi_2|,
\qquad
1-|\xi|^2 = \lambda\,(\widetilde{\Delta E\,\tau})^2 ,
\end{equation}
without any fine tuning of parameters: the bipartite structure protects the interference. Second, the general phase difference \eqref{eq:sm_Delta_theta_general} collapses to
\begin{equation}\label{eq:sm_Delta_theta}
\Delta\theta =
-(E_{12}-E_{34})\tau
+ 2\eta\,\widetilde{\Delta E\,\tau},
\end{equation}
reproducing Eqs.~\eqref{eq:leading_theta} and \eqref{eq:Fn_freq}. 

Finally, substituting $1-|\xi|^2 = \lambda(\widetilde{\Delta E\tau})^2$ and $|\xi|^{2n} = [1-\lambda(\widetilde{\Delta E\tau})^2]^n \simeq e^{-\lambda(\widetilde{\Delta E\tau})^2 n}$ into the interference formula \eqref{eq:sm_interference_final} yields
\begin{equation}
    F_n \sim
    4 \left[ \lambda (\widetilde{\Delta E \tau})^2 \right]^{2} e^{-\lambda (\widetilde{\Delta E \tau})^2 n} \cos^2 \left( \frac{n}{2} \Delta \theta + \beta \right),
\end{equation}
which is Eq.~\eqref{eq:Fn_spectral}. The detuning thus controls the lifetime of the interference through the envelope $e^{-\lambda(\widetilde{\Delta E\tau})^2 n}$, while the frequency remains finite at resonance, $\Delta\theta\to-(E_{12}-E_{34})\tau\neq0$: the two ``temporal slits'' stay separated.

% \subsection{Merging of three phase factors: derivation of Eq.~(10)}
\subsection{Merging of three phase factors \texorpdfstring{[Eqs.~\eqref{eq:Fn_triple},~\eqref{eq:theta_slow}]}{}}\label{sec:sm_case2}

The second scenario of the main text arises in bipartite systems with an odd number of energy levels, where a zero-energy state exists. Shifting the energies so that the central level is at $E=0$ (an overall energy shift only produces a global phase), the resonance involves the symmetric triplet $\{-E,\,0,\,+E\}$ with weights $\{P,\,P_0,\,P\}$, where $P_0\equiv P_{(E=0)}$ and $P\equiv P_{(E)}$ in the notation of the main text.
%; the symmetric placement of the outer levels and the equality of their weights follow from the bipartite symmetry (Sec.~\ref{sec:sm_bipartite}). 
At resonance, $e^{-iE\tau}=e^{+iE\tau}=1$ and, following the mapping to the charge picture of Appendix~\ref{app:charge} and Ref.~\cite{yin2019}, the three charges coalesce at the point $1$ of the unit circle. Slightly away from resonance, with detuning $\varepsilon \equiv \widetilde{\Delta E\,\tau}$, $\Delta E = E-0$, the charges sit at $1$ and $e^{\mp i\varepsilon}$.

The perturbative result for the magnitudes now reads \cite{yin2019}
\begin{equation}\label{eq:sm_triple_modulus}
|\xi_{1,2}|^2 \simeq 1-\lambda'\,(\widetilde{\Delta E\,\tau})^2,
\qquad
\lambda' = \frac{P}{(P_0+2P)^2},
\end{equation}
which is the damping factor quoted below Eq.~\eqref{eq:Fn_triple}. 
%Both dominant eigenvalues have the same magnitude at this order: here the equal-decay condition is enforced by the reflection symmetry of the triplet itself.

For the phases we again treat the merging group in isolation. Normalizing the strengths of the three charges as $P_0' = P_0/(P_0+2P)$ and $P' = P/(P_0+2P)$, so that $P_0'+2P'=1$, the force balance reads
\begin{equation}\label{eq:sm_triple_balance}
  \frac{P_0'}{1-\xi} + P'\left(\frac{1}{e^{-i\varepsilon}-\xi} + \frac{1}{e^{i\varepsilon}-\xi}\right) = 0.
\end{equation}
Exploiting the small parameter $\varepsilon$, after straightforward algebra this reduces to the quadratic equation
\begin{equation}
  \begin{aligned}
  &\xi^2 - B\,\xi + C = 0,\\
  &B = 2P_0'\cos\varepsilon + 2P'(1+\cos\varepsilon) = 2 - \frac{1+P_0'}{2}\,\varepsilon^2 ,\\
  &C = P_0' + 2P'\cos\varepsilon = 1 - \frac{1-P_0'}{2}\,\varepsilon^2 ,
  \end{aligned}
\end{equation}
up to $\mathcal{O}(\varepsilon^4)$. The two solutions are
\begin{equation}\label{eq:sm_triple_phases}
  \xi_{1,2} \simeq 1 \mp i \sqrt{P_0'}\,\varepsilon
  \;\;\Rightarrow\;\;
  \theta_{1,2} \simeq \mp\,\sigma\, \widetilde{\Delta E\,\tau},
\end{equation}
with $\sigma = \sqrt{P_0'} = \sqrt{P_0/(P_0+2P)}$ the geometric factor defined below Eq.~\eqref{eq:theta_slow},
where we label $\xi_1$ as the eigenvalue pulled toward the charge $e^{-iE\tau}$ ($E>0$). Hence
\begin{equation}
\Delta\theta_{\rm slow} = \theta_1-\theta_2 = -2\sigma\,\widetilde{\Delta E\,\tau},
\end{equation}
which is Eq.~\eqref{eq:theta_slow}: in contrast to the two-pair scenario, the leading term in Eq.~\eqref{eq:leading_theta} vanishes identically, and the oscillation frequency itself is linear in the detuning, vanishing at the exact resonance where the two eigenvalues coalesce. Substituting Eq.~\eqref{eq:sm_triple_modulus} into Eq.~\eqref{eq:sm_interference_final} yields Eq.~\eqref{eq:Fn_triple},
\begin{equation}
    F_n \sim 4 \left[ \lambda' (\widetilde{\Delta E \tau})^2 \right]^{2} e^{-\lambda' (\widetilde{\Delta E \tau})^2 n} \cos^2 \left( \frac{n}{2} \Delta \theta_{\text{slow}} + \beta \right).
\end{equation}
Since $\Delta\theta_{\rm slow}\propto\widetilde{\Delta E\,\tau}$ while the envelope decays $\propto(\widetilde{\Delta E\tau})^2$, close to resonance many measurements fit within a fraction of an oscillation period, producing the wide plateau of nearly frozen detection probability shown in Fig.~\ref{fig:Fn_spectral}(c).

%We have verified all the formulas of this section against the exact eigenvalues of $\mathcal{S}$; representative numbers for the models of the Letter are collected in Table~\ref{tab:sm_parameters} of Sec.~\ref{sec:sm_parameters}.

\section{Bipartite symmetry and structural protection of the interference}\label{sec:sm_bipartite}

A graph $G$ is \emph{bipartite} if its vertices can be partitioned into two disjoint sublattices $A$ and $B$ such that every edge connects vertices of different sublattices. In the $(A,B)$ basis the tight-binding Hamiltonian reads
\begin{equation}
H = -\gamma\!\sum_{u\in A,\,v\in B}\big(\ket{u}\bra{v}+\ket{v}\bra{u}\big)
= \begin{pmatrix} 0 & T\\ T^\dagger & 0 \end{pmatrix},
\end{equation}
where $\gamma$ is the hopping rate and $T$ is the $N_A\times N_B$ coupling matrix between the sublattices, with $N_A$ and $N_B$ the numbers of vertices in $A$ and $B$. With the %chiral (sublattice)
sublattice operator
\begin{equation}
\Gamma=\begin{pmatrix} \mathbb{1}_A & 0\\ 0 & -\mathbb{1}_B \end{pmatrix},
\end{equation}
the block-off-diagonal form of $H$ gives the exact bipartite %(chiral)
symmetry \cite{spectra_book}
\begin{equation}\label{eq:sm_bipartite_symmetry}
\Gamma H \Gamma = -H .
\end{equation}
Two well-known consequences follow.

\textit{Spectral pairing.---}If $H\ket{E}=E\ket{E}$ with $E\neq0$, then Eq.~\eqref{eq:sm_bipartite_symmetry} gives $H\,\Gamma\ket{E}=-\Gamma H\ket{E}=-E\,\Gamma\ket{E}$, so $\Gamma\ket{E}$ is an eigenstate of energy $-E$. The spectrum is therefore symmetric about zero, with eigenvalues in $\pm E$ pairs (plus possible zero modes). At a sampling time $\tau$ the two partners give a complex-conjugate pair of charges on the unit circle, $e^{-iE\tau}$ and $e^{+iE\tau}=(e^{-iE\tau})^*$, as visible in the eigenvalue distribution of Fig.~\ref{fig:Fn_4hypercube}(b).

\textit{Equal weights.---}If the target state is localized on a single sublattice, $\Gamma\ket{x_T}=\pm\ket{x_T}$, then $|\langle x_T|\Gamma|E\rangle|^2 = |\langle x_T|E\rangle|^2$, and since $\Gamma$ maps the $E$ eigenspace onto the $-E$ eigenspace,
\begin{equation}\label{eq:sm_equal_P}
P_{+E}=P_{-E},
\end{equation}
so the two partner charges carry equal strength. In particular $P_3=P_1$ and $P_4=P_2$ for the conjugate resonant pairs of Appendix~\ref{sec:sm_case1}, and the outer weights of the triplet of Appendix~\ref{sec:sm_case2} are equal. Since a single vertex always lies in one sublattice, this condition holds automatically for the recurrence problems of this work.

What are the consequences for the eigenvalues of $\mathcal{S}$? Combining the two, the charge configuration $\{(e^{-iE_k\tau},P_k)\}$ on the unit circle is invariant under complex conjugation. The electrostatic field of Eq.~\eqref{eq:force_field},
\begin{equation}
\mathcal{F}(\xi)=\sum_{k=1}^w \frac{P_k}{e^{-iE_k\tau}-\xi},
\end{equation}
whose stationary points are the eigenvalues of $\mathcal{S}$, then obeys $[\mathcal{F}(\xi^*)]^*=\mathcal{F}(\xi)$. Hence if $\xi$ is an eigenvalue so is $\xi^*$: the eigenvalues of $\mathcal{S}$ are either real or occur in complex-conjugate pairs, symmetric about the real axis, which is fixed by the zero-energy center of the bipartite spectrum. In particular, near a $w\to w-2$ resonance built from two conjugate pairs (or a symmetric triplet), the two dominant eigenvalues form a conjugate pair,
\begin{equation}\label{eq:sm_conjugate_pair}
\xi_2=\xi_1^*,\qquad \theta_2=-\theta_1,\qquad |\xi_1|=|\xi_2|,
\end{equation}
consistent with the perturbative results of Appendix~\ref{sec:sm_perturbation} and manifest in Figs.~\ref{fig:Fn_4hypercube}(b) and \ref{fig:Fn_spectral}(b,d).

\section{Spectral theory of $F_n$ for the four-dimensional hypercube}\label{sec:sm_hypercube_spectral}

In the main text the spectral prediction, Eq.~\eqref{eq:Fn_spectral}, for the merging of two pairs of phase factors is illustrated on the seven-site tight-binding chain [Fig.~\ref{fig:Fn_spectral}(a)]. Here we show that the same theory quantitatively captures the monitored quantum walk on the four-dimensional hypercube, the model of Figs.~\ref{fig:fig_protocol1} and \ref{fig:Fn_4hypercube}.

The hypercube Hamiltonian [Eq.~\eqref{eq:hypercube_ham}] is bipartite, with distinct energy levels $E\in\{-4,-2,0,2,4\}$ and weights $P=\{1,4,6,4,1\}/16$, so the generic winding number is $w=5$. At $\tau=\pi/3$ the two conjugate pairs $(E_1,E_2)=(4,-2)$ and $(E_3,E_4)=(-4,2)$ merge simultaneously, driving the $w=5\to3$ transition marked in Fig.~\ref{fig:Fn_4hypercube}(a). This is precisely the two-pair scenario of Appendix~\ref{sec:sm_case1}: the bipartite symmetry enforces $P_3=P_1$ and $P_4=P_2$, so the two merging pairs carry matched weights, and the two dominant eigenvalues of $\mathcal{S}$ approach the unit circle together.

\begin{figure}[h]
    \centering
    \includegraphics[width=0.48\textwidth]{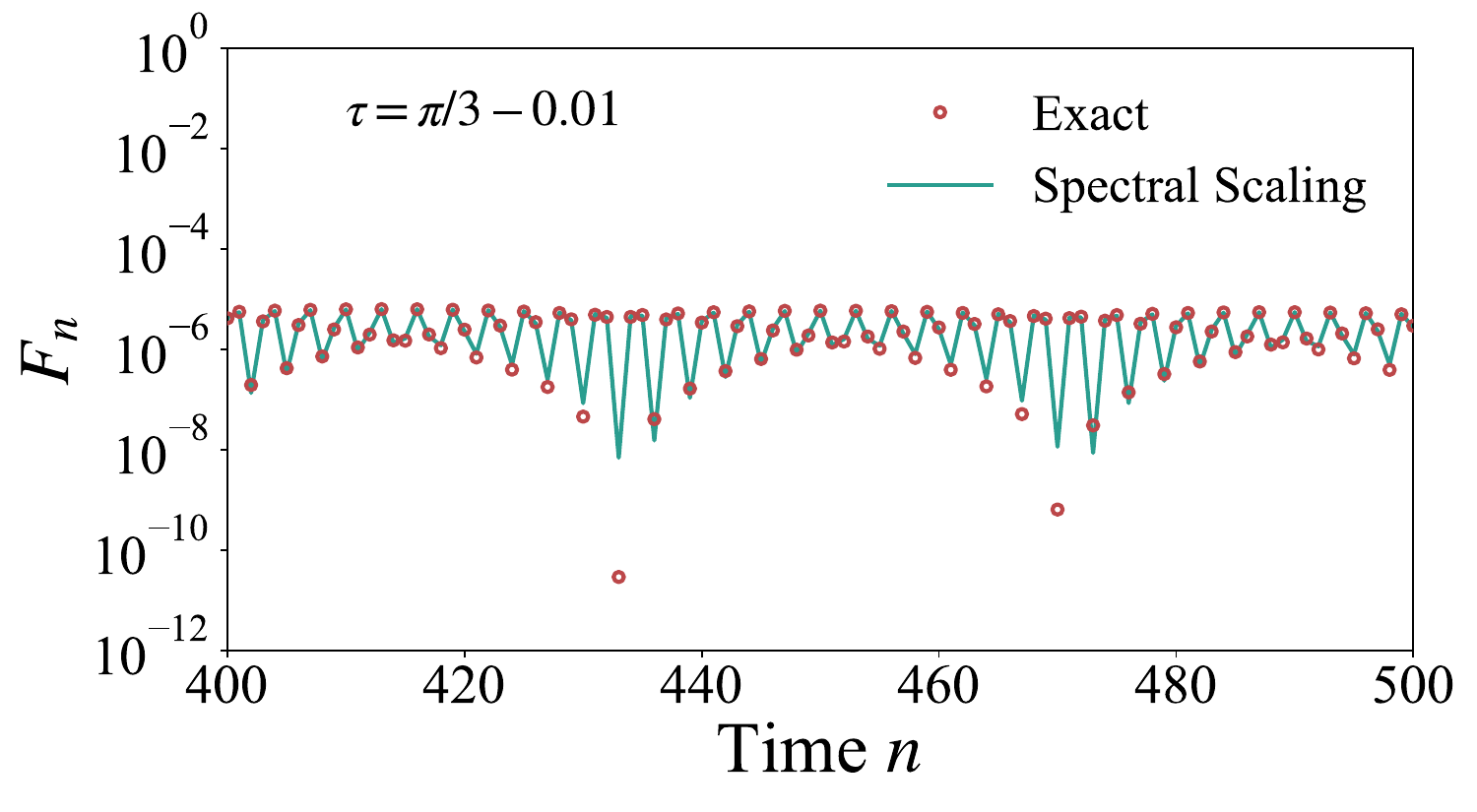}
    \caption{First-detection probability $F_n$ versus measurement number $n$ for the monitored quantum walk on the four-dimensional hypercube at $\tau=\pi/3-0.01$, close to the $w=5\to3$ transition. The exact result from Eq.~\eqref{eq:phi_n} (red circles) agrees with the spectral prediction of Eq.~\eqref{eq:Fn_spectral} (solid line), obtained from the merging of two pairs of phase factors (Appendix~\ref{sec:sm_case1}) with $\lambda=64/125$, $\eta=-3/10$, and detuning $\widetilde{\Delta E\,\tau}=-0.06$. The two-pair mechanism reproduces both the slow envelope decay and the finite-frequency oscillation.}
    \label{fig:sm_hypercube_spectral}
\end{figure}

Tuning slightly off resonance to $\tau=\pi/3-0.01$, the detuning becomes $\widetilde{\Delta E\,\tau}=-0.06$, and Eqs.~\eqref{eq:Fn_spectral} and \eqref{eq:Fn_freq} apply with $\lambda = P_1P_2/(P_1+P_2)^3 = 64/125$ and $\eta=(P_1-P_2)/[2(P_1+P_2)]=-3/10$ (using $P_1=1/16$ and $P_2=4/16$). Figure~\ref{fig:sm_hypercube_spectral} compares the exact $F_n$ from Eq.~\eqref{eq:phi_n} (red circles) with the spectral prediction of Eq.~\eqref{eq:Fn_spectral} (solid line) at long times. The excellent agreement confirms that the two-pair interference mechanism---slow envelope decay $e^{-\lambda(\widetilde{\Delta E\tau})^2 n}$ together with a finite oscillation frequency $\Delta\theta$---carries over unchanged from the chain to the hypercube, and, more generally, applies to any bipartite system.

The merging of three phase factors (Appendix~\ref{sec:sm_case2}), by contrast, requires an isolated conjugate pair $\pm E$ to coalesce with the zero-energy level while all other phase factors stay well separated. In the four-dimensional hypercube the nonzero energies $\pm2$ and $\pm4$ are commensurate, so
no sampling time isolates a single triplet at the zero level. Bringing the $\pm2$ pair onto the zero level ($\tau=\pi$) draws the $\pm4$ pair along with it, joining all four nonzero levels to the zero level at once---a fivefold coalescence, $w\to w-4$. Bringing the $\pm4$ pair onto the zero level ($\tau=\pi/2$) instead leaves the $\pm2$ pair to coalesce with itself at phase factor $e^{\pm i \pi}=-1$, opposite to the $+1$ shared by the merged triplet $\{0,\pm4\}$, producing this triplet alongside a separate pair $\{\pm2\}$, i.e.\ $w\to w-3$. In neither case is an isolated triplet realized. The isolated triplet is instead realized in hypercubes of other dimensions; for example, the two-dimensional hypercube has spectrum $\{-2,0,2\}$ with weights $\{1,2,1\}/4$, precisely the symmetric triplet of Appendix~\ref{sec:sm_case2}.

\section{Parameters for the examples of \texorpdfstring{the main text}{the main text}}\label{sec:sm_parameters}

We list the resonance structure and the numerical parameters of the perturbative theory (Appendix~\ref{sec:sm_perturbation}) for the two models shown in the figures of the main text. The models are defined in Appendices~\ref{app:hypercube} and \ref{app:chain}, Eqs.~\eqref{eq:hypercube_ham} and \eqref{eq:chain_ham}, with $\gamma=1$.

\emph{Hypercube, $d=4$ [Figs.~\ref{fig:fig_protocol1} and \ref{fig:Fn_4hypercube}].} The distinct energy levels are $E\in\{-4,-2,0,2,4\}$ with weights $P=\{1,4,6,4,1\}/16$, so the generic winding number is $w=5$. As detailed in Appendix~\ref{sec:sm_hypercube_spectral}, at $\tau=\pi/3$ the two conjugate pairs $(4,-2)$ and $(-4,2)$ merge, producing the $w=5\to3$ transition of Fig.~\ref{fig:Fn_4hypercube}(a); the figures use the detuned time $\tau=\pi/3-0.01$, with detuning $\widetilde{\Delta E\,\tau}=-0.06$ and merging-pair weights $P_1=1/16$, $P_2=4/16$, giving $\lambda=64/125$ and $\eta=-3/10$. At $\tau=\pi/4$, in contrast, only the single pair $(4,-4)$ merges, giving the $w=5\to4$ transition and the monotonic decay of Appendix~\ref{app:monotonic}.

\emph{Chain, $L=7$ [Fig.~\ref{fig:Fn_spectral}].} The energy levels are $E_k = -2\cos(k\pi/8)$, $k=1,\dots,7$, with weights $P_k = \tfrac14 \sin^2(k\pi/8)$, and $w=7$ generically. Two resonances are used in Fig.~\ref{fig:Fn_spectral}:
(i)~at $\tau = 2\pi/(\sqrt{2}+\sqrt{2+\sqrt{2}})$ the conjugate pairs $(E_1,E_2)=(\sqrt{2+\sqrt2},\,-\sqrt2)$ and $(E_3,E_4)=(-\sqrt{2+\sqrt2},\,\sqrt2)$ merge, which is the two-pair scenario of Appendix~\ref{sec:sm_case1}, with weights $P_1=(2-\sqrt2)/16$ and $P_2=1/8$, giving $\lambda=16(19+3\sqrt2)/343$ and $\eta=-(1+2\sqrt2)/14$;
(ii)~at $\tau=\sqrt2\,\pi$ the phase factors of the triplet $\{-\sqrt2,\,0,\,\sqrt2\}$ coalesce at unity, which is the three-charge scenario of Appendix~\ref{sec:sm_case2}, with $P_0 = 1/4$ and $P=1/8$, giving $\lambda' = 1/2$ and $\sigma = 1/\sqrt2$.

Table~\ref{tab:sm_parameters} compares the perturbative predictions with the exact values obtained from the eigenvalues of $\mathcal{S}$ at the detuned sampling times used in the figures of the main text. The predicted frequency, Eq.~\eqref{eq:sm_Delta_theta}, agrees closely with the exact values.
\begin{table*}[h]
  \centering
  \caption{Perturbation theory versus exact values for the examples of the main text. For each case we quote the resonant sampling time $\tau_R$, the sampling time $\tau$ used in the figures, the detuning $\widetilde{\Delta E\,\tau}$, the envelope parameter $1-|\xi|^2$ [theory: $\lambda(\widetilde{\Delta E \tau})^2$ or $\lambda'(\widetilde{\Delta E \tau})^2$], and the oscillation frequency $|\Delta\theta|$ [theory: $\Delta\theta=-(E_{12}-E_{34})\tau+2\eta\,\widetilde{\Delta E\,\tau}$ for the two-pair case, $\Delta\theta_{\rm slow}=-2\sigma\,\widetilde{\Delta E\,\tau}$ for the triple case]. Exact values follow from the two dominant eigenvalues of $\mathcal{S}$, Eq.~\eqref{eq:app_eigen}.}
  \label{tab:sm_parameters}
  \setlength{\tabcolsep}{4pt}%
  \renewcommand{\arraystretch}{1.25}%
  \small
  \begin{tabularx}{\textwidth}{>{\raggedright\arraybackslash}Xccccc}
  \hline\hline
  Case & $\tau_R$ & $\tau$ & $\widetilde{\Delta E\,\tau}$ & $(1-|\xi|^2)\times10^{3}$: theory / exact & $|\Delta\theta|$: theory / exact \\
  \hline
  Hypercube $d{=}4$, two pairs [Figs.~\ref{fig:fig_protocol1}, \ref{fig:Fn_4hypercube}] & $\pi/3$ & $\pi/3-0.01$ & $-0.0600$ & $1.843$ / $1.832$ & $2.0384$ / $2.0382$ \\
  Chain $L{=}7$, two pairs [Fig.~\ref{fig:Fn_spectral}(a,b)] & $2\pi/(\sqrt2+\sqrt{2+\sqrt2}\,)$ & $\tau_R-0.02$ & $-0.0652$ & $4.615$ / $4.735$ & $0.7907$ / $0.7910$ \\
  Chain $L{=}7$, triple [Fig.~\ref{fig:Fn_spectral}(c,d)] & $\sqrt{2}\,\pi$ & $\tau_R - 0.02$ & $-0.0283$ & $0.400$ / $0.400$ & $0.0400$ / $0.0400$ \\
  \hline\hline
  \end{tabularx}
\end{table*}

\end{document}